\documentclass[12pt]{article}
\pdfoutput=1
\usepackage{amssymb,amsmath}
\usepackage{epsfig}
\usepackage{epstopdf}
\usepackage{latexsym}
\usepackage{graphicx}
\usepackage{subfigure}
\usepackage{booktabs}
\usepackage{bbm}
\usepackage[margin=20pt,small]{caption}

\usepackage[toc]{appendix}

\usepackage{color}
\usepackage{datetime}

\ifpdf
\fi

\renewcommand{\theequation}{\arabic{section}.\arabic{equation}}

\def\cM{{\cal M}}
\def\cN{{\cal N}}
\def\cO{{\cal O}}

\def\cL{{\cal L}}

\def\cL{{\cal L}}

\def\cphi{{\Delta \phi_{*}}}

\definecolor{cardinal}{rgb}{0.6,0,0}
\definecolor{darkgreen}{rgb}{0,0.5,0}
\definecolor{golden}{rgb}{0.92, 0.7, 0}
\definecolor{midnight}{rgb}{0, 0, 0.5}
\definecolor{darkblue}{rgb}{0.2, 0, 0.8}

\begin{document}

\begin{titlepage}

\begin{flushright}
DIAS-STP-12-06\\
UTTG-10-12 \\
TCC-013-12
\end{flushright}

\bigskip
\centerline{\Large \bf Back-reaction of Non-supersymmetric Probes:}
\bigskip
\centerline{\Large \bf Phase Transition and Stability}
\bigskip
\bigskip
\centerline{{\bf Matthias Ihl$^{1}$, Arnab Kundu$^{2}$ and Sandipan Kundu$^{2,3}$}}
\bigskip
\centerline{$^1$ School of Theoretical Physics}
\centerline{Dublin Institute for Advanced Studies} 
\centerline{10 Burlington Rd, Dublin 4, Ireland.}
\bigskip
\centerline{$^2$ Theory Group, Department of Physics}
\centerline{University of Texas at Austin} 
\centerline{Austin, TX 78712, USA.}
\bigskip
\centerline{$^3$ Texas Cosmology Center}
\centerline{University of Texas at Austin} 
\centerline{Austin, TX 78712, USA.}
\bigskip
\centerline{msihl@stp.dias.ie, arnab@physics.utexas.edu, sandyk@physics.utexas.edu}
\bigskip
\bigskip

\begin{abstract}

\noindent {We consider back-reaction by non-supersymmetric D7/anti-D7 probe branes in the Kuperstein-Sonnenschein model at finite temperature. Using the smearing technique, we obtain an analytical solution for the back-reacted background to leading order in $N_f/N_c$. This back-reaction explicitly breaks the conformal invariance and introduces a dimension $6$ operator in the dual field theory which is an irrelevant deformation of the original conformal field theory. We further probe this back-reacted background by introducing an additional set of probe brane/anti-brane. This additional probe sector undergoes a chiral phase transition at finite temperature, which is absent when the back-reaction vanishes. We investigate the corresponding phase diagram and the thermodynamics associated with this phase transition. We also argue that additional probes do not suffer from any instability caused by the back-reaction, which suggests that this system is stable beyond the probe limit.} 

\end{abstract}

\newpage

\tableofcontents

\end{titlepage}

\newpage

\section{Introduction}

The understanding of the gauge-gravity duality or the Anti--de-Sitter/Conformal Field Theory (AdS/CFT) correspondence \cite{Maldacena:1997re, Gubser:1998bc, Witten:1998qj} (see {\it e.g.}~\cite{Aharony:1999ti} for an earlier review) has provided us with a large class of strongly coupled large $N_c$ gauge theories. The very nature of this duality enables us to perform computations in classical general relativity which --- when translated to the gauge theory language {\it via} the AdS/CFT dictionary --- corresponds to a computation in the dual field theory in the strong coupling limit. This is a remarkably powerful technique to probe aspects of strongly coupled gauge theories where conventional perturbative field theory methods are severely inadequate.

Strongly coupled systems are abundant in nature: the physics of the strong interaction described by Quantum Chromodynamics (QCD), as suggested by its name, describes a strongly coupled system. Coincidentally we live in an interesting time when the strong-coupling physics of QCD has been experimentally explored in the Relativistic Heavy Ion Collider (RHIC) and is being currently explored in the Large Hadron Collider (LHC); see {\it e.g.}~\cite{Adcox:2004mh, Arsene:2004fa, Back:2004je, Adams:2005dq, Aamodt:2010pa} for more information about experimental results and findings. These explorations and findings necessitate improved theoretical control in addressing strong coupling physics.

This is where the AdS/CFT correspondence makes her\footnote{We apologize if this gender assignment appears presumptuous. It seems the age-old folklore of a conventional assignment to the explorer and the explored has gotten the better of us.} entrance. Currently we do not have a gravitational dual to QCD in a precise sense. However, within this class of large $N_c$ gauge theories, the strongly coupled dynamics of the ``quarks" and the ``gluons" can be explored and understood. Such theories are typically supersymmetric and conformal with adjoint matter fields ($\equiv$ ``gluons"), and in addition fundamental matter ($\equiv$ ``quarks") can be introduced. The hope is that by studying these theories, we can learn general lessons about universal features of strongly coupled systems that also apply --- at least qualitatively --- to QCD.

In this article, our focus will be both on the fundamental flavor sector and its back-reaction on the adjoint sector. Typically the fundamental matter fields are introduced by considering $N_f$ number of flavor branes in the background of $N_c$ color branes. The stack of color branes gives rise to a gravitational background which can be obtained by solving the supergravity equations of motion. In this background the flavor branes are introduced in the probe limit, with $N_f \ll N_c$, such that they do not back-react on the background. This idea was pioneered in \cite{Karch:2002sh}. Many interesting physical questions can be explored and understood in the flavor sector within this probe limit and within various models, {\it e.g.}~the meson spectrum and the meson melting phase transition respectively in \cite{Kruczenski:2003be} and \cite{Mateos:2006nu, Albash:2006ew, Karch:2006bv} in the D3-D7 model. Another canonical example is the physics of the chiral symmetry breaking within the flavor sector realized in {\it e.g.}~the so called Sakai-Sugimoto model \cite{Sakai:2004cn, Sakai:2005yt} and its finite temperature version in \cite{Aharony:2006da, Parnachev:2006dn};\footnote{It is argued in \cite{Mandal:2011ws} that the chiral symmetry restoration mechanism proposed in \cite{Aharony:2006da, Parnachev:2006dn} is problematic and a new mechanism is proposed. We thank Takeshi Morita for pointing this out to us.} and more recent models of chiral symmetry breaking in \cite{Kuperstein:2008cq, Dymarsky:2009cm}.\footnote{For a comparative study of the D3-D7 model and the Sakai-Sugimoto model in the presence of various external parameters, see {\it e.g.}~\cite{Kundu:2010ye}.}

All these analyses are performed within the probe limit where it is assumed that the number of flavors is very small compared to the number of colors. In QCD, however, this is not the case. First we have $N_c=3$ and thus large $N_c$ gauge theories will not work as a good quantitative approximation and we may only hope for robust qualitative features. Second, we have $N_f \sim N_c$; hence the physics that we observe in the strict probe limit may also be of limiting use. One may wonder how much the qualitative physics can really depend on the precise values of $N_f$ and $N_c$ and their relative strength. It certainly depends on the precise physical question, however we can identify certain physical properties that crucially depend on these values. One such example is the QCD beta function: it can be shown that QCD is an asymptotically free theory as long as $N_f < \frac{11}{2} N_c$, which holds true in nature. However, the physics will be completely different if the above inequality is not satisfied. Another crucial property is chiral symmetry breaking where the physics of the chiral transition depends on $N_c$ and $N_f$ leading to a phase structure which is sensitive to these numbers (at least on their ballpark ranges).\footnote{See {\it e.g.}~\cite{Iwasaki:1996ny} for some lattice results.} Thus we can safely conclude that going beyond the probe limit, as far as introducing flavor degrees of freedom is concerned, is a rather important topic to explore.\footnote{This limit is known as the Veneziano limit.}

The general procedure to consider back-reaction consists of the following steps: We start with a background obtained by solving the supergravity equations of motion, {\it e.g.}~an AdS$_5\times M_5$ background, where $M_5$ is some Sasaki-Einstein manifold. This gravitational background is obtained from the closed string sector. The dual field theory is some $\cN=1$ superconformal quiver gauge theory with a certain global symmetry group which is identified with the isometry group of the internal manifold. To introduce flavors, we need to add open string degrees of freedom. The dynamics of the open string degrees of freedom is described by the probe D-brane action of appropriate dimension and is given by the so-called Dirac-Born-Infeld (DBI) action (supplemented by the Wess-Zumino term whenever necessary). To consider the back-reaction we need to consider the DBI action as the matter source that deforms the original background.

Work along this direction has been undertaken by several authors over the years beginning with {\it e.g.}~\cite{Bigazzi:2005md, Casero:2006pt, Paredes:2006wb, Benini:2006hh, Benini:2007gx, Caceres:2007mu} and nicely summarized in \cite{Nunez:2010sf}. These references have analyzed the back-reaction by supersymmetric probes at zero temperature. Supersymmetry makes the problem more tractable since one does not need to solve the equations of motion, but instead it is sufficient to solve first order BPS equations.  This technical simplification comes at the cost that the resulting theory --- even after the inclusion of the flavor degrees of freedom --- is supersymmetric and does not realize the spontaneous breaking of chiral symmetry as one would expect in QCD.\footnote{Note that the issue of the back-reaction in the Sakai-Sugimoto model, where both supersymmetry is completely broken and we have a geometric realization of spontaneous chiral symmetry breaking, has been addressed in \cite{Burrington:2007qd}.} Nonetheless, numerous interesting features have been explored within this type of models at zero and finite temperature and more recently including other parameters, such as a chemical potential or an external electromagnetic field in {\it e.g.}~\cite{Bigazzi:2009bk, Bigazzi:2011it, Bigazzi:2011db, Magana:2012kh, Filev:2011mt, Ammon:2012qs}.

Here we will focus on a model recently discussed in \cite{Kuperstein:2008cq} where D7 and anti-D7 branes have been introduced in the AdS$_5\times T^{1,1}$ background. The brane--anti-brane pair wraps a three cycle in the internal manifold $T^{1,1} \cong S^2 \times S^3$ and extends along the rest of the cone. At zero temperature, the brane--anti-brane pair joins in the IR realizing a spontaneous breaking of the chiral symmetry: $U(N_f)_L \times U(N_f)_R \to U(N_f)_{\rm diag}$. The asymptotic angle separation between the brane and the anti-brane --- previously denoted by $\Delta\phi_\infty$, but will henceforth be denoted by $\cphi$ for reasons that will become apparent in the main text --- takes a constant value and corresponds to the coupling of the corresponding operator in the flavor sector. In \cite{Alam:2012fw} we have discussed in details the phase structure of this model in the presence of a non-zero temperature and an external electro-magnetic field.

Before going further, let us comment on the relevance of the Kuperstein-Sonnenschein model. First, even at zero temperature, the probes break supersymmetry completely and realize a spontaneous breaking of chiral symmetry. Furthermore, unlike the Sakai-Sugimoto model, the dual field theory is an honest $(3+1)$-dimensional one. Thus, this model potentially realizes certain properties of QCD better than the Sakai-Sugimoto model. In this article, we will consider the back-reaction by the flavor sector in this model at finite temperature.

The rationale behind considering back-reaction in the finite temperature case is three-fold: first note that since the probes break supersymmetry completely, no simplification such as solving first order BPS equations will occur at zero temperature. Second, the profile at zero temperature is described by a non-trivial function $\phi(r)$ (where $\phi$ is the azimuthal angle of the $S^2$ where the probes are extended), which makes analyzing the back-reaction problem more involved.\footnote{Recall that when the background is AdS-Schwarzschild$\times T^{1,1}$, the favorable embeddings are described by $\phi ={\rm const}$ as shown in \cite{Alam:2012fw}.} Third, for our current purposes --- which is to analyze the phase structure once the back-reaction is taken into account --- the relevant background should have a non-zero temperature.

The astute reader might raise the following question: how is this system any different from what has been obtained in {\it e.g.}~\cite{Bigazzi:2009bk} which also deals with a finite temperature back-reacted solution? This is indeed a very relevant question. For now we will offer a brief answer to this question through some technicalities involved in the computations and elaborate on these throughout the rest of the paper. First recall that the precise embedding as considered in {\it e.g.}~\cite{Casero:2006pt} and the ones that we consider here {\it a la} \cite{Kuperstein:2008cq} are different in the details. These details turn out to be important for the physics that we eventually extract from such back-reacted systems. Second, in our system, we have both D7 and anti-D7 branes as opposed to having just branes in the systems considered so far. The D7-branes are codimension two objects and therefore they act as two dimensional delta function sources. The resulting equations of motion become coupled PDEs which pose a challenge. This difficulty can be circumvented by using the so-called ``smearing" technique which means we introduce many such probes and distribute them in the transverse directions homogeneously. If we have an equal number of branes and anti-branes smeared at each point, it will not source any $C_8$ potential (the Hodge dual to the axion field). This is in stark contrast to the framework considered in {\it e.g.}~\cite{Bigazzi:2009bk} where there are no anti-branes and therefore a non-trivial $C_8$ potential is sourced.

Complying with the conventional wisdom, we find that the flavor back-reaction is understood as irrelevant deformations of the original CFT. In our case, these deformations consist of the departure from conformality and the insertion of a dimension $6$ operator. The former originates from a running dilaton and the later from the deformation of both the $S^2 \times S^2$ base and the $U(1)$ fiber direction of the original $T^{1,1}$-manifold. We will not make attempts to precisely identify this dimension $6$ operator, however we will extract a lot of information about the ``phase structure" of this theory in a similar spirit as in \cite{Alam:2012fw}.

The breaking of the conformal invariance due to flavor back-reaction has a very important physical consequence. To appreciate this, let us recall that in the original Klebanov-Witten background the flavor sector did not undergo a chiral phase transition at finite temperature simply because there was no other scale available \cite{Alam:2012fw}. The back-reaction of the flavors introduces this dimension $6$ operator, equivalently a dimension $-2$ coupling denoted by $\mathfrak{m}^{-2}$, providing another natural scale in the system. We can now probe this back-reacted background with an additional set of $N_f'$ D7/anti-D7 branes and focus on the physics this extra set of flavor sector experiences. The interplay of the background temperature and the dimension $-2$ coupling conspires to conjure up a non-trivial phase structure. We have studied this phase structure in the $\cphi$ vs $m^2$ plane, where $m^2$ is defined as
\begin{eqnarray}
m^2 =  \pi L^4 \left(\frac{T^2}{\mathfrak{m}^2}\right) \left(1 + \frac{\epsilon}{8} \right) \ , \quad \epsilon = \frac{3}{2\pi^2} \left(\frac{\lambda N_f}{N_c}\right) \ ,
\end{eqnarray}
Here $T$ is the background temperature, $L$ is the radius of the original AdS space and $\lambda$ is the 't Hooft coupling. 

Another important and interesting question regarding such non-supersymmetric probing is the issue of stability: it is not entirely clear how stability might be affected once the back-reaction by the non-supersymmetric probes are taken into account. It was demonstrated in \cite{Dymarsky:2010ci} that the non-supersymmetric probes are stable in the Kuperstein-Sonnenschein model. Here we will not undertake the task of analyzing the fluctuation modes of the back-reacted background. We will, however, analyze the stability of the additional set of probes that we introduce in this back-reacted background. We will argue that this additional probe sector does not experience any unbalanced force due to the absence of supersymmetry and the fluctuations of these additional probes are not likely to experience any instability caused by the back-reaction itself. Thus this is also very suggestive of the fact that perhaps stability is not a matter of concern in this model in which supersymmetry plays no role. We will comment more on this issue in the concluding part of this work.

This paper is organized as follows: we start with a brief review of the Kuperstein-Sonnenschein model at finite temperature in section 2. In section 3, we discuss the issue of taking back-reaction into account, describe the analytical solution treating the back-reaction perturbatively and offer comments on the dual field theory. In the following section, we discuss various thermodynamic properties of this back-reacted background. Then, in section 5, we introduce an additional set of probe brane/anti-brane system in this back-reacted background and study the phase structure in the probe sector. In section 6, we argue that this probe sector does not suffer from any instability after the inclusion of the back-reaction. Finally we conclude in section 7 with future directions. Various technical details have been relegated to four appendices: appendix A contains a discussion of the smearing form, appendix B contains the details of the equations of motion, in appendix C we have provided the detailed steps towards obtaining the leading order back-reacted solution and in appendix D we have obtained the action corresponding to the quadratic fluctuations of the additional probe sector in our back-reacted background.

\section{Kuperstein-Sonnenschein probes at finite temperature}

In \cite{Kuperstein:2008cq}, the authors introduced probe D7/anti-D7 branes in the AdS$_5 \times T^{1,1}$ background such that the D7/anti-D7 branes wrap a three cycle in the internal manifold $T^{1,1} \cong S^2 \times S^3$ and are extended along the rest of the conifold $\mathbb{R}^2 \times S^2$. The field theory dual to this supergravity background is given by an $\cN =1$ superconformal quiver gauge theory with a gauge group $SU(N_c) \times SU(N_c)$ and two bi-fundamental chiral superfields which are usually denoted by $A_{1,2}$ and $B_{1,2}$. These chiral superfields transform in the $\left(N_c, \bar{N_c}\right)$ and the $\left(\bar{N_c}, N_c\right)$ representations of the gauge group. This theory also possesses a global $SU(2) \times SU(2) \times U(1)_R$ symmetry which, in the dual gravitational background, manifests itself as the isometry group of the internal manifold $T^{1,1}$.

Introducing probe D7/anti-D7 branes amounts to introducing matter fields in the fundamental representation: specifically it corresponds to adding left-handed/right-handed Weyl fermions in the dual gauge theory. This gives rise to a global symmetry group of $U(N_f)_L \times U(N_f)_R$ flavor symmetry, where $N_f$ is the number of flavors that we have introduced in this background.

\begin{figure}[h!]
\begin{center}
\subfigure[] {\includegraphics[angle=0,
width=0.45\textwidth]{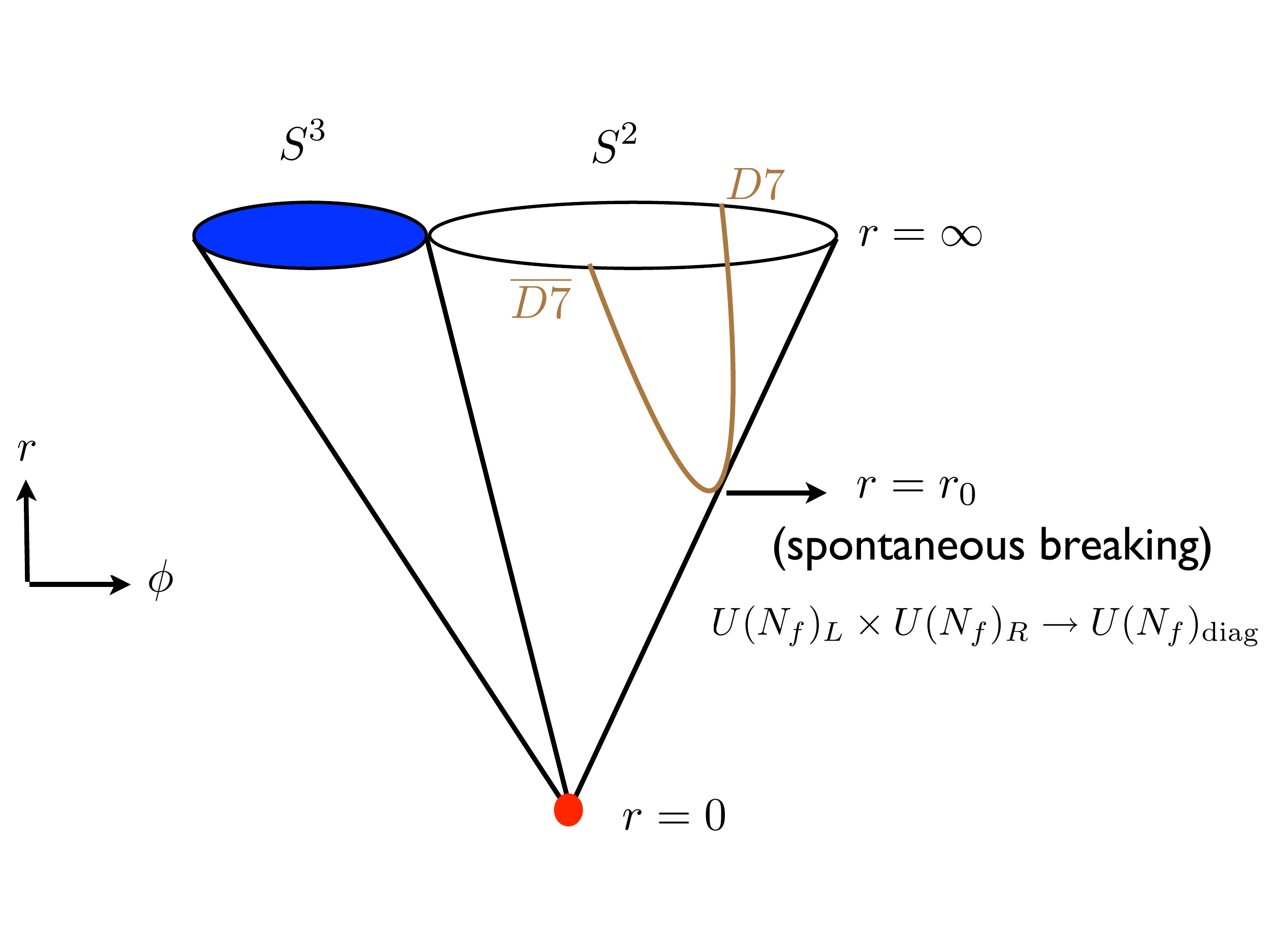} }
 \subfigure[] {\includegraphics[angle=0,
width=0.45\textwidth]{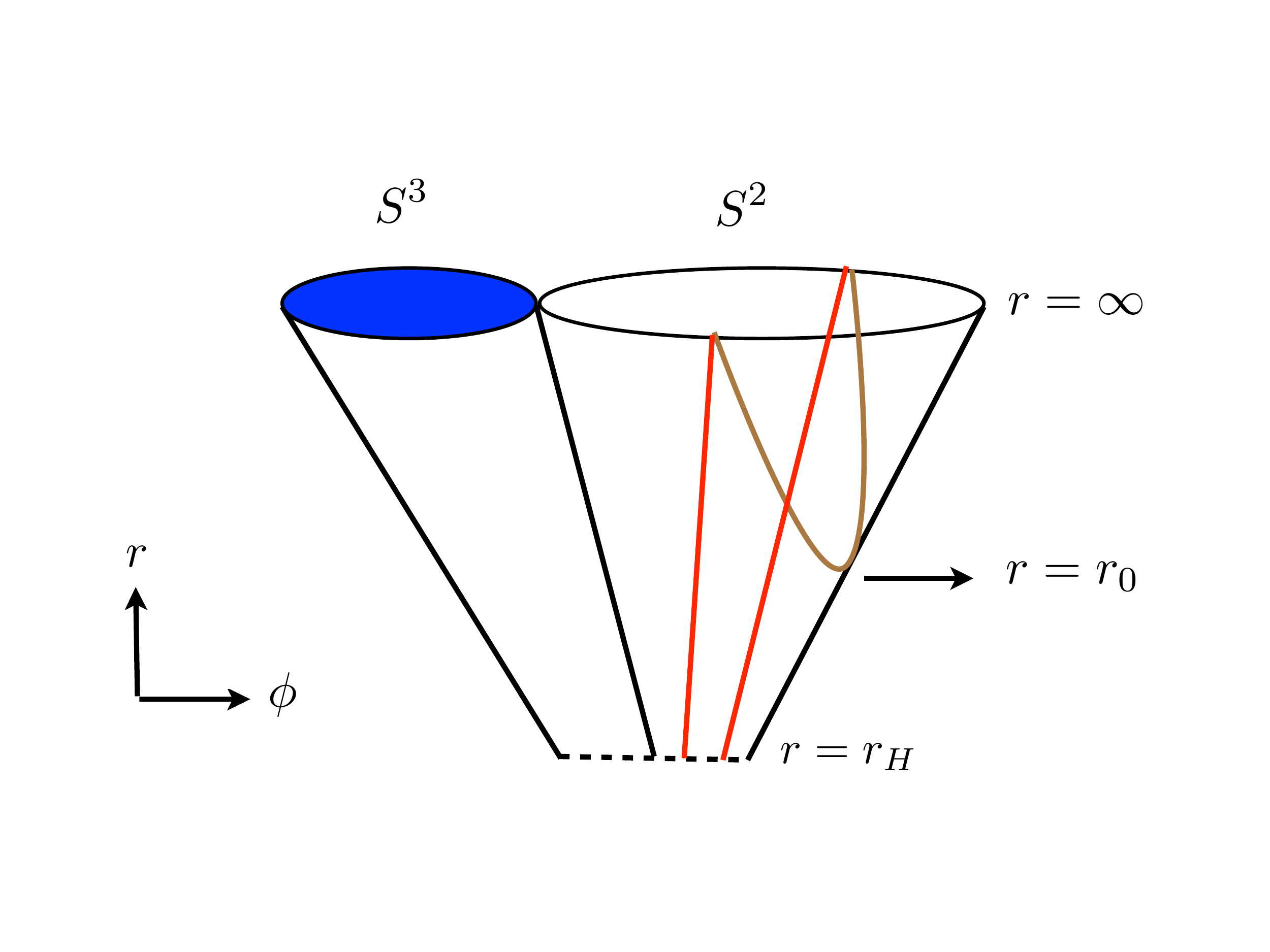} }\caption{A schematic representation of the shapes of the probe branes. Fig.~(a) The red dot denotes the conifold singularity located at $r=0$. The probe branes join at $r=r_0$ realizing the spontaneous breaking of chiral symmetry. Fig.~(b): Two types of embedding are possible at finite temperature and the only available phase is the chiral symmetry restored phase.}
\label{kwprobe}
\end{center}
\end{figure}

At zero temperature, the probe D7 and anti-D7 branes have no choice but to dynamically join in the IR and realize spontaneous breaking of chiral symmetry: $U(N_f)_L \times U(N_f)_R \to U(N_f)_{\rm diag}$. In the presence of a black hole horizon in the bulk (and therefore finite temperature in the dual field theory), there are two types of profile: one where the brane--anti-brane pair joins at some IR radial position; and the other where they separately fall into the black hole. The former corresponds to a chiral symmetry broken phase and the later to the chiral symmetry restored phase. This phenomenon is pictorially represented in fig.~\ref{kwprobe}.

Although we have two different types of embeddings, there is no possibility of a phase transition between them. The reason is simple: since our background is conformal, temperature is the only scale. Thus there is no meaning of having a ``high" or a ``low" temperature phase, there will be an unique phase at any temperature. As argued and explicitly shown in \cite{Alam:2012fw}, introducing a non-zero temperature immediately favors the chiral symmetry restored phase.

Let us be more specific. The background at finite temperature is given by AdS$_5$-Schwarzschild$\times T^{1,1}$.\footnote{In this case the dilaton is a constant, therefore the string frame metric and the Einstein frame metric are the same up to a factor of the string coupling $g_s$. In keeping with the notations used in the later parts of the paper, we will define $g_s := e^{\Phi_*}$.} The time direction is Euclideanized and compacitified on a circle. The temperature is then simply given by the inverse period of this circle. In Lorentzian signature this background is given by
\begin{eqnarray}
&& ds^2 = \frac{r^2}{L^2} \left( - b(r) dt^2 + d\vec{x}^2 \right) + \frac{L^2}{r^2} \frac{dr^2}{f(r)} + R^2 ds_{T^{1,1}}^2 \ , \quad b(r) = 1- \frac{r_H^4}{r^4} \ , \\
&& ds_{T^{1,1}}^2 = \frac{1}{6} \left[ \left( e^1\right)^2 + \left( e^2\right)^2 + \left( \tilde{e}^1\right)^2 + \left( \tilde{e}^2\right)^2 \right] + \frac{1}{9} \left( e^3 + \tilde{e}^3\right)^2 \ .
\end{eqnarray}
Here $t$ is the time, $\vec{x}$ represents the spatial $3$-directions, $r_H$ is the location of the event-horizon and $L$ is the radius of curvature of the AdS-space. After Euclideanization, the temperature is given by $T = r_H / (\pi L^2)$. Also $e^i$ and $\tilde{e}^i$ are the left-invariant one-forms of two independent $SU(2)$'s. There is an $U(1)$ symmetry which rotates $e^3$ to $\tilde{e}^3$. Finally there is a $\mathbb{Z}_2$ symmetry which acts as $\mathbb{Z}_2 : e^{1,2} \leftrightarrow \tilde{e}^{1,2}$. Thus the background has an $SU(2) \times SU(2) \times U(1) \times \mathbb{Z}_2$ symmetry, where the $U(1)$ is the R-symmetry.

We choose the vielbeins as follows\cite{Papadopoulos:2000gj}
\begin{eqnarray}
&& e^1 = \sin\frac{\psi}{2} \sin\theta_1 d\phi_1 + \cos\frac{\psi}{2} d\theta_1 \ , \quad e^2 =  - \cos\frac{\psi}{2} \sin\theta_1 d\phi_1 + \sin\frac{\psi}{2} d\theta_1 \ , \\
&& \tilde{e}^1 = \sin\frac{\psi}{2} \sin\theta_2 d\phi_2 + \cos\frac{\psi}{2} d\theta_2 \ , \quad \tilde{e}^2 =  - \cos\frac{\psi}{2} \sin\theta_2 d\phi_2 + \sin\frac{\psi}{2} d\theta_2 \ , \\
&& e^3 + \tilde{e}^3 = d\psi + \cos\theta_1 d\phi_1 + \cos\theta_2 d\phi_2 \ .
\end{eqnarray}
Here $\{\theta_1, \phi_1\}$ and $\{\theta_2, \phi_2 \}$ parametrize the two $S^2$s. The ranges for the angular directions are given by: $\psi \in [0, 4\pi)$, $\phi_i \in [0, 2\pi)$ and $\theta_i \in [0, \pi)$. With this parametrization, the metric on the $T^{1,1}$ becomes
\begin{eqnarray}
ds_{T^{1,1}}^2 & = & \frac{1}{6} \sum_{i=1,2}\left( d\theta_i^2 + \sin^2\theta_i d\phi_i^2 \right) + \frac{1}{9} \left( d\psi + \sum_{i=1,2}\cos\theta_i d\phi_i \right)^2 \nonumber\\
& = & ds_{\rm KE}^2 + \left( d\psi + A_{\rm KE} \right)^2 \ ,
\end{eqnarray}
where in the second line the metric on $T^{1,1}$ has been written as the $U(1)$ fibration over a K\"{a}hler-Einstein manifold $S^2 \times S^2$ and $A_{\rm KE}$ yields the K\"{a}hler form {\it via} $J = d A_{\rm KE} /2$.

The probe branes are aligned such that they span the Minkowski- and $\{\psi, \theta_{1,2}, \phi_{1,2}\}$-directions and their profile is represented by $\{\theta_{2,1}(r), \phi_{2,1}(r)\}$. The DBI action in this case is given by
\begin{eqnarray} \label{probepure}
S_{\rm DBI} = - \tau_7 \int d^8 \xi \sqrt{{\rm det} P[G]} = - \cN \int dt dr r^3 \left( 1 + \frac{r^2}{6} b(r) \left(\theta_{1,2}'^2 + \sin^2\theta_{1,2} \, \phi_{1,2}'^2 \right)\right)^{1/2} \ ,
\end{eqnarray}
where $\tau_7$ is the tension of the probe brane and $\cN = (\tau_7 V_{\mathbb{R}^3} 8 \pi^2) / 9$. This action leads to equations of motion which admit two different classes of solutions represented by
\begin{eqnarray}
&& \theta_{1,2} = \pi/2 \ , \quad \phi_{1,2} = \phi(r) \ , \\
&& \theta_{1,2} = \pi/2 \ , \quad \phi_{1,2} = {\rm const} \ .
\end{eqnarray}
The profiles described by a non-constant $\phi(r)$ are called the U-shaped embeddings and they correspond to the chiral symmetry broken phase; the other ones are called parallel-shaped and correspond to the chiral symmetry restored phase. These two classes of profiles are schematically shown in fig.~\ref{kwprobe}. It was shown in \cite{Alam:2012fw} that energetics always favors the chiral symmetry restored phase in the presence of a non-zero temperature. In the subsequent sections we will analyze the back-reaction on the supergravity background by these parallel-shaped brane/anti-brane profiles and obtain the back-reacted background at finite temperature.

\section{The back-reacted background}

From here on we will discuss the effect of back-reaction by the flavor branes at finite temperature. The probes in our system are co-dimension two objects and thus they give rise to delta function sources in the plane transverse to the seven branes. Thus, the general problem of considering back-reaction by the flavor branes will lead to a set of coupled partial differential equations which is difficult to solve. To simplify this problem we can make use of the so-called ``smearing" technique as demonstrated in {\it e.g.}~\cite{Benini:2006hh}.

The smearing procedure can be understood as uniformly distributing the large number of probes along the transverse directions such that the global symmetry of the original background is recovered after taking the back-reaction into account. Following the discussion in \cite{Benini:2006hh}, we briefly elaborate on this idea. The smearing process for $N_f$ D7-branes can be operationally understood as the following modification to the Wess-Zumino contribution
\begin{eqnarray}
S_{\rm WZ} = \tau_7 \sum^{N_f} \int_{\cM_8} C_8 \to \tau_7 \int_{\cM_{10}} \Omega \wedge C_8 \ ,
\end{eqnarray}
where $\Omega$ is a two-form known as the smearing form,\footnote{We will find this smearing form in our case is simply given by a constant. Further details on this computation have been relegated in appendix A.} $\cM_{8}$ denotes the D7-brane world-volume and $\cM_{10}$ denotes the full ten-dimensional manifold. This two-form determines the distribution of the RR-charges sourced by the D7-branes along the transverse directions. Clearly, if we have an equal number of uniformly smeared D7 and anti-D7 branes the Wess-Zumino contributions will cancel each other which implies that there is no net RR-charge at each point along the directions where the smearing procedure has been implemented.\footnote{Thus the back-reaction in our system will not source any $C_8$ (the dual to the axion field) potential and consequently we will not have any $F_1$ RR-form. This is to be contrasted with the case in {\it e.g.}~\cite{Benini:2006hh} where the back-reaction sources a non-trivial axion field.}

Moreover, due to the smearing procedure the DBI-piece of the probe action --- in Einstein frame --- will take the form
\begin{eqnarray}
S_{\rm DBI} = - \tau_7 \int_{\cM_8} d^8\xi e^{\Phi} \sqrt{- G_8} \to - \tau_7 \int_{\cM_{10}} d^{10}x e^{\Phi} \sqrt{- G_{10}} \left | \Omega \right | \ ,
\end{eqnarray}

So in view of the above discussion, to consider the back-reaction we will work with the following action in Einstein frame
\begin{eqnarray} \label{acback}
S & = & S_{\rm SUGRA} + S_{\rm flavour} \nonumber\\
& = & \frac{1}{2\kappa_{10}^2} \int d^{10}x \, \sqrt{-G_{10}} 
\Big[ R - \frac{1}{2} \partial_M \Phi \partial^M \Phi -\frac{1}{4} |F_5|^2 \Big]  \nonumber\\
& - &  2 \tau_7 \sum^{N_f} \int d^{8}\xi \, e^\Phi \Big[ 
\sqrt{-G_{8}^{(1)}} \Big]  -   2 \tau_7 \sum^{N_f} \int d^{8}\xi \, e^\Phi \Big[ 
\sqrt{-G_{8}^{(2)}} \Big]  \ , \nonumber\\
\end{eqnarray}
where the superscript on the induced metric $G_8^{(1,2)}$ is intended to denote the probes which are located at $\theta_{1,2} = \pi/2$ and $\phi_{1,2} = {\rm const}$. Also, $\kappa_{10}$ is related to the ten-dimensional Newton's constant, $R$ is the Ricci scalar, $G_{10}$ is the background metric, $\Phi$ is the dilaton and $F_5$ is the self-dual five-form field strength. The supergravity part of the action scales as $g_s^{-2}$ and the flavor part of the action scales as $g_s^{-1}$. This is because the supergravity part emerges from the closed string sector whereas the flavor part originates from the open string sector. Note that we have defined the dilaton field such that $\Phi$ vanishes asymptotically and the constant $\Phi_*$, which sets the string coupling {\it via} the relation $g_s^* \equiv g_s : = e^{\Phi_*}$, has been factored out.\footnote{Here we will use the symbol $g_s^*$ to denote the string coupling in the presence of a non-zero back-reaction in keeping with the other notations.} The quantity $\tau_7$ is the tension of the brane: $\tau_7 = g_s^{-1} (2\pi)^{-7} \alpha'^4$. The factor of $2$ in front of the DBI-action comes from the brane and the anti-brane which contribute equally. More details on the action and the corresponding equations of motion have been relegated to appendix B.

Now that we know which action to extremize to find the desired solution, we should start with an ansatz for the back-reacted background. The full ten-dimensional metric should be a deformation of the AdS$_5$-Schwarzschild$\times T^{1,1}$ background. More precisely --- since the D7-branes wrap the entire AdS part --- the deformation should appear only within the internal $T^{1,1}$-part. Recall that the smearing procedure has recovered the full $SU(2) \times SU(2) \times U(1) \times {\mathbb Z}_2$ global symmetry. Thus the internal manifold is still constraint to possess this isometry and --- much like the case considered in \cite{Benini:2006hh} --- we make the following ansatz
\begin{align} \label{ansatz}
ds_{10}^2 &= h^{-1/2}(r)\left[-b(r) dt^2 + dx_i dx^i\right]+
h^{1/2}(r) \bigg\{\frac{dr^2}{b(r)} +
\frac{e^{2g(r)}}{6} \sum_{i=1,2} \left( d\theta_i^2 + \sin^2 \theta_i \, d\phi_i^2 \right)  \nonumber \\
& + \frac{e^{2f(r)}}{9} \left(d\psi + \sum_{i=1,2} \cos\theta_i \, d\phi_i \right)^2 \bigg\} \ , \\
F_5 &= k(r)h(r)^{3/4}\left(e^t \wedge e^{x^1} \wedge e^{x^2} \wedge e^{x^3} \wedge e^r + e^{\psi} \wedge e^{\theta_1} \wedge e^{\phi_1} \wedge e^{\theta_2} \wedge e^{\phi_2}\right) \ ,
\end{align}
where $h(r)$, $b(r)$, $f(r)$ and $g(r)$ are unknown functions and the vielbeins are defined as
\begin{eqnarray}
&& e^t = h^{-1/4} b^{1/2} dt \ , \quad e^{x^i} =  h^{-1/4} dx^i \ , \quad e^r = h^{1/4} b^{-1/2} dr \ , \\
&& e^{\psi} = \frac{1}{3} h^{1/4} e^f \left( d\psi + \cos\theta_1 d\phi_1 + \cos\theta_2 d \phi_2 \right)  \ , \\
&& e^{\theta_{1,2}} =  \frac{1}{\sqrt{6}} h^{1/4} e^g d\theta_{1,2} \ ,  \quad e^{\phi_{1,2}} = \frac{1}{\sqrt{6}} h^{1/4} e^g \sin\theta_{1,2} d\phi_{1,2} \ .
\end{eqnarray}
We also have a non-constant dilaton $\Phi(r)$. The quantization condition for the five form flux
\begin{eqnarray}
\frac{1}{2\kappa_{10}^2} \int F_{5}=N_c \tau_3 \ ,
\end{eqnarray}
where $\tau_3$ is the tension of the D3-branes, gives
\begin{eqnarray}
&& k(r) h(r)^2 e^{4g(r)+f(r)}=27 \pi g_s^* N_c l_s^4  = 4 L^4 : = \frac{27}{4} \lambda l_s^4 \ , \\
&& {\rm with} \quad \lambda = 4 \pi g_s^* N_c  \ . \label{lambda}
\end{eqnarray}
With the ansatz in (\ref{ansatz}) one can now obtain six independent equations for the five unknown functions obtained by extremizing the action in (\ref{acback}). These equations are shown and discussed in details in appendix B. We will not discuss these details here, instead we will move on to the analytical solution that can be obtained for the leading order correction in $N_f/N_c$ in the next section.

\subsection{Perturbative solution}

We will now exhibit the perturbative solution that we obtained treating the back-reaction as a ``small" effect. Before discussing this solution explicitly, let us be precise about what we mean by this ``smallness". From the action we can explicitly check that the back-reaction part of the action (which is given by the DBI-piece) is suppressed by a factor of $N_f/g_s \sim (\lambda N_f)/N_c$ as compared to the supergravity action. This is expected since the supergravity action scales as $N_c^2$ and the probe action in (\ref{probepure}) scales as $(\lambda N_f N_c)$. Thus to obtain a perturbative solution the relevant expansion parameter, denoted by $\epsilon$, will be $\epsilon \sim (\lambda N_f)/N_c$.

To fix the numerical coefficient, we look at the right hand side of {\it e.g.}~equation (\ref{eqns}) and define
\begin{eqnarray} \label{eps}
\epsilon := \frac{6}{\pi} g_s^* N_f = \frac{3}{2\pi^2} \left(\frac{\lambda N_f}{N_c}\right) \ ,
\end{eqnarray}
where we have used the definition of $\lambda$ from equation (\ref{lambda}). We will discuss the back-reacted solution at the leading order in $\epsilon$.

In fact it turns out that we can find an analytical solution in the leading order in $\epsilon$: in appendix C we have discussed the details on how to obtain this solution. Here we will only present the final result and discuss its properties. The various functions are obtained to be
\begin{eqnarray}
&& b(r) =\left(1-\frac{r_H^4}{r^4}\right) \ ,\label{solb} \\
&&  \Phi(r)=\frac{\epsilon}{4}\ln\left(\frac{r}{r_*}\right) \ , \label{solPhi}\\
&& h(r)=\frac{L^4}{r^4}\left(1+\frac{\epsilon}{8}\right) \ ,\\
&& e^{f(r)}=r\left[1+\epsilon\left(-\frac{1}{24}+4 \mathfrak{m}^{-2}  \left(\frac{r_H^4}{r^2}\right)K\left(1-\frac{r_H^4}{r^4}\right)- 8 \mathfrak{m}^{-2} r^2 E\left(1-\frac{r_H^4}{r^4}\right)\right)\right] \ , \label{solf} \\
&& e^{g(r)}=r\left[1+\epsilon\left(-\frac{1}{48}- \mathfrak{m}^{-2}  \left(\frac{r_H^4}{r^2}\right)K\left(1-\frac{r_H^4}{r^4}\right)+2 \mathfrak{m}^{-2} r^2 E\left(1-\frac{r_H^4}{r^4}\right)\right)\right] \ . \label{solg}
\end{eqnarray}
Here $r_H$ is the location of the event-horizon, $L$ is the radius of the AdS-space, $K$ and $E$ are elliptic functions of the first and the second kind respectively and $\mathfrak{m}^{-2}$ is an undetermined constant of dimension ${\rm length}^{-2}$. Thus we have actually found a one parameter family of solutions parametrized by this constant. Also, $r_*$ is a UV cut-off. The dilaton runs logarithmically and thus blows up at $r \to \infty$, which implies that the string coupling also blows up. Thus we need to define this theory with a UV cut-off at $r = r_*$ where $\Phi \to 0$.\footnote{Recall that we have already factored out the string coupling from the dilaton and therefore the dilaton has to approach zero asymptotically.}

Let us now discuss some properties of this solution. Note that the solution for the emblackening factor $b(r)$ and the warp factor $h(r)$ do not receive any qualitative correction due to the back-reaction. At most we can redefine the AdS-radius $L$ by absorbing the $\epsilon$ correction to $h(r )$, which will define an effective 't Hooft coupling for us. On the other hand, the functions $e^f$ and $e^g$ do receive interesting corrections. At first it might seem that the back-reaction changes the asymptotic behavior of these functions, but by the very nature of being a perturbative solution this is not the case. We will discuss the regime of validity of this solution in the next section where it will become clearer.

It is evident from the perturbative solution given in (\ref{solb})-(\ref{solg}) that the flavor back-reaction deforms the original AdS$_5 \times T^{1,1}$-space and this deformation shows up entirely in the internal manifold $T^{1,1}$. Such deformations are understood {\it via} the GKPW formula\cite{Gubser:1998bc, Witten:1998qj} by looking at the asymptotic behavior of the fields
\begin{eqnarray}
\delta X = a_X r^{\Delta - 4} + v_X r^{-\Delta} \ ,
\end{eqnarray}
where $\delta X$ denotes the deformation of the metric component, $a_X$ and $v_X$ are two constants and $\Delta$ is the dimension of the corresponding operator. In the dual theory, such a deformation corresponds to deforming the Lagrangian by $\cL_{\rm deform} = \cL_{\rm CFT} + a_X \cO$, where $\cO$ denotes the corresponding operator and $v_X = \langle \cO \rangle$ denotes the vacuum expectation value (VEV) of the field.

From our solution, especially using the expressions in (\ref{solf}) and (\ref{solg}), it can be shown that the deformed metric functions have the following asymptotic behaviors
\begin{eqnarray}
h^{1/2}\{e^{2f} , e^{2g} \} \sim  \left[ 1 + \epsilon \, \beta_{f,g} \,  \mathfrak{m}^{-2} r_*^2 \right]   \ ,
\end{eqnarray}
where $\beta_{f,g}$ is some dimensionless numerical constant that will not be important for us, and $r_*$ is the UV cut-off. It is clear from the above expansion that $\mathfrak{m}^{-2}$ corresponds to a coupling which has dimension $-2$ in the dual field theory. Thus the deformation to the CFT in this case will be of the form: $\cL_{\rm deform} = \cL_{\rm CFT} + \mathfrak{m}^{-2} \cO$, where ${\rm dim}[\cO] = 6$. The non-normalizable term in the above expansion of the metric function means we have inserted an irrelevant operator of dimension $6$ in the dual field theory. This fact conforms to the conventional wisdom that flavor back-reaction generally gives rise to irrelevant deformations of the background\cite{Benini:2006hh}.

Let us be more precise about the identification of the various fields in the gravity background with the operators in the dual field theory.\footnote{We are thankful to Javier Tarrio for pointing this out to us.} Following the discussion in section 3.4 of \cite{Benini:2006hh}, the scalar fields that are dual to the dimension $6$ and the dimension $8$ operators --- which are denoted respectively by $p$ and $q$ --- are given by 
\begin{eqnarray}
p = - \frac{1}{5} \left( f - g \right) \ , \quad q = \frac{2}{15}\left( f + 4 g + \frac{5}{4} \log h \right) \ .
\end{eqnarray}
Now using the most general solution that is presented in (\ref{solgenb})-(\ref{solgeng}), we get
\begin{eqnarray}
p =  \epsilon \frac{2}{\mathfrak{m}^2} r_*^2  + \ldots \ ,  \quad q = \frac{2}{3} \log L + \frac{\epsilon}{3} \left(\frac{c_3}{L^4} + \frac{2 c_5}{5 r_H^4} \right) r_*^4  + \ldots \ .
\end{eqnarray}
The choice made in (\ref{c3c5choice}) is equivalent to setting the source for the dimension $8$ operator to zero.

Note that, as far as the gravitational background is considered, the deformations are very much similar to the ones discussed in {\it e.g.}~\cite{Benini:2006hh}: {\it i.e.}~because of the running dilaton, we have a deformation from the conformal Klebanov-Witten solution to the non-conformal one. There is also a relative metric deformation between the $S^2 \times S^2$ base and the $U(1)$ fiber direction of the original $T^{1,1}$. As is demonstrated in {\it e.g.}~\cite{Gubser:1998kv}, the departure from the conformal invariance is sourced by ${\rm Tr} F^4$-term in the dual CFT, which is a dimension $8$ operator. As for the identification of the dimension $6$ operator (discussed in the previous paragraph) is concerned, we will remain agnostic since the Lagrangian of the probe sector itself is not well-understood at present. However, this will not prevent us from exploring the phase diagram of this theory as the coupling $\mathfrak{m}^{-2}$ is varied.\footnote{Perhaps a better expression would be ``the family of theories", since varying a coupling  by hand means we are continually changing the theory.}

\subsection{A few comments on the regime of validity}

To comment on the regime of validity of the perturbative solution, let us begin by recalling the solution for the dilaton
\begin{eqnarray}
\Phi =  \frac{\epsilon}{4} \log \left(\frac{r}{r_*}\right)  \ .
\end{eqnarray}
Quite clearly, the dilaton blows up exactly in the limit $r \to \infty$. Thus the dilaton in our first order back-reacted background diverges only at infinity, which means there exists a Landau pole when the UV is located strictly at infinity.\footnote{We will elaborate on this point in the next section.} This is similar to the situation studied in \cite{Bigazzi:2009bk} where the Landau pole appears at $r\to\infty$ if we are to take the leading order solution in $\epsilon$. Thus, as far as the Landau pole is concerned, we can safely use our solutions for $r_* \ll \infty$, where $r_*$ is the UV cut-off.

Now to make sense of the perturbative solution given in (\ref{solb})-(\ref{solg}) for the range of radial coordinate $r_H \le r \le r_*$ we must impose
\begin{eqnarray} 
&& \epsilon \left | \log\left(\frac{r_H}{r_*}\right) \right | \ll \cO(1)  \quad {\rm and} \label{valid1} \\
&& \epsilon \left | \mathfrak{m}^{-2} \left( \frac{r_H^4}{r_*^2}\right) \right | \ll 1 \ , \quad \epsilon \left | \mathfrak{m}^{-2} r_*^2  \right | \ll 1 \label{valid2}
\end{eqnarray}
along with the condition that $\epsilon \ll 1$. The first condition comes from demanding that the perturbative solution for the dilaton makes sense and the second two conditions come from demanding the same for the functions $e^f$ and $e^g$.

As discussed in \cite{Bigazzi:2009gu}, such back-reacted solutions can have other UV pathologies. Such pathologies show up in the so-called ``holographic $a$-function". It can also be checked explicitly that the holographic $a$-function in our case always remains finite and there is no additional UV cut-off in the game. This is most likely because we  found a solution only up to the leading order in the back-reaction. Thus we have a simple hierarchy of scales in our system
\begin{eqnarray}
r_H \ll r_* \ll r_{\rm LP} \ , \quad {\rm with} \quad r_{\rm LP} \to \infty \ ,
\end{eqnarray}
where the subscript ``LP" stands for Landau pole.

The validity of the SUGRA+DBI etc will give identical conditions as in \cite{Bigazzi:2009bk}. Suppression of closed string loops imposes $N_c \gg 1$ and the suppression of $\alpha'$ corrections imposes $\lambda \gg 1$. For the smearing procedure we also need a large number of probes, which means $N_f \gg 1$. We will hold $N_f/ N_c$ fixed and this is essentially the parameter that we control. Also, we need to impose that $\alpha'$ corrections to the supergravity background (which comes from leading order terms like $\alpha'^3 R^4$ and hence scales as $\lambda^{-3/2}$) are sub-leading compared to the flavor back-reaction. This further imposes that $\lambda^{-3/2} \ll \epsilon$.

\subsection{A few comments on the dual field theory}

In this section we will offer some comments regarding the dual field theory. We will closely follow the discussion presented in \cite{Benini:2006hh}. Before taking the back-reaction into account, the field theory dual to the AdS$_5\times T^{1,1}$ is an $\cN=1$ superconformal quiver gauge theory with gauge group $SU(N_c)\times SU(N_c)$. This theory has two bi-fundamental matter fields, usually denoted by $A_i$ and $B_i$, $i=1,2$ with a quartic superpotential\cite{Klebanov:1998hh} 
\begin{eqnarray}
W = \lambda {\rm Tr} \left( A_i B_j A_k B_l \right) \epsilon^{ik} \epsilon^{jl} \ ,
\end{eqnarray}
where $\lambda$ is the 't Hooft coupling. This theory has a global $SU(2)\times SU(2) \times U(1)_R \times \mathbb{Z}_2$ symmetry.

Adding the probe sector {\it a la} \cite{Kuperstein:2008cq} introduces chiral flavors and realizes a global $U(N_f)_L \times U(N_f)_R$ flavor symmetry, where $N_f$ is the number of D7 and anti-D7 branes. At zero temperature this flavor symmetry is spontaneously broken down to a $U(N_f)_{\rm diag}$\cite{Kuperstein:2008cq}, but at finite temperature this symmetry is restored\cite{Alam:2012fw}. Since this theory is not supersymmetric by construction, it is not currently clear how the Lagrangian of the probe sector should look like. For some thoughts regarding this issue, see \cite{Kuperstein:2008cq}. Nonetheless, a lot of interesting physics can be extracted purely using the gauge-gravity duality.

When considering the back-reaction, we are modifying the theory. The general procedure of taking into account back-reaction by the flavor sector introduces localized sources and therefore the global symmetry is expected to be of the form $U(N_f)_L\times U(N_f)_R$. However, to avoid the complexity of the problem and to have good control we have taken advantage of the ``smearing procedure" as outlined earlier and in appendix A. The smearing procedure distributes the branes and the anti-branes such that we recover the full global symmetry of the internal manifold that we had before introducing any flavors. The smearing is done homogeneously such that there is one D7 and one anti-D7 at each point of the two $S^2$'s. Thus the flavor symmetry group is broken down to two copies of 
\begin{eqnarray}
U(N_f)_L\times U(N_f)_R \to U(1)_L^{N_f} \times U(1)_R^{N_f} \ .
\end{eqnarray}

Now, we are considering the back-reaction at finite temperature. The presence of a non-zero temperature breaks the conformal symmetry in the sense that it introduces a scale in the system. In the probe limit, the flavor sector breaks this conformal symmetry spontaneously even at zero temperature\cite{Kuperstein:2008cq}. Thus the inclusion of the back-reaction will explicitly break conformal invariance and will lead to a non-zero beta function. In order to make this precise, we need to know the relations between the gauge couplings, the theta angles and the supergravity fields. Such relations can be derived properly for the $\cN = 2$ orbifold theory which is dual to the background obtained by placing a large number of D3-branes at the singularity of $\mathbb{C} \times \mathbb{C}^2/\mathbb{Z}_2$. In this orbifold theory such relations are given by
\begin{eqnarray}
&& \frac{4\pi^2}{g_1^2} + \frac{4\pi^2}{g_2^2} = \frac{\pi e^{-\Phi}}{g_s} \ , \\
&&  \frac{4\pi^2}{g_1^2} - \frac{4\pi^2}{g_2^2} = \frac{e^{-\Phi}}{g_s} \left[ \frac{1}{2\pi\alpha'} \int_{S^2} B_2 - \pi \quad \left( {\rm mod} \, \, 2\pi \right)\right] \ ,  \\
&& \theta_1^{\rm YM} = - \pi C_0 + \frac{1}{2\pi}\int_{S^2} C_2 \quad \left( {\rm mod} \, \, 2\pi \right) \ , \\
&& \theta_2^{\rm YM} = - \pi C_0 - \frac{1}{2\pi}\int_{S^2} C_2 \quad \left( {\rm mod} \, \, 2\pi \right) \ ,
\end{eqnarray}
where $g_{1,2}$ and $\theta_{1,2}$ are the two gauge couplings and theta angles corresponding to the two gauge groups, $B_2$ is the NS-NS two-form, $C_2$ is the RR two-form and $C_0$ is the axion field. Here the integrals are performed over the two-sphere which shrink at the orbifold fixed point. We will assume that the above relations hold true for the conifold theory as well.\footnote{However, in \cite{Strassler:2005qs} it has been suggested that the formulae relating the sum of the gauge couplings and the theta angles should be modified. Nonetheless, in the spirit of \cite{Benini:2006hh} we will use these formulae to extract qualitative features.}

In our back-reacted background, we do not have any $H_3 = dB_2$, $C_2$ or $C_0$. Assuming that the two gauge couplings are equal $(g_{\rm YM}^2:= g_1^2/2 = g_2^2/2)$, we get
\begin{eqnarray}
\frac{4\pi^2}{g_{\rm YM}^2} = \frac{\pi}{g_s^*} e^{- \Phi} \ .
\end{eqnarray}
Using the definition of the 't Hooft coupling $\lambda = g_{\rm YM}^2 N_c$, the solution for the dilaton from equation (\ref{solPhi}), the definition of $\epsilon$ from (\ref{eps}) and the above relation we can compute the beta function to be given by
\begin{eqnarray}
r \beta_{\lambda} = r \left(\partial_r \lambda \right)= \left(\frac{3}{8\pi^2}\right) \frac{N_f}{N_c} \lambda^2 > 0 \ .
\end{eqnarray}
Thus the beta function acquires a positive contribution of the order $N_f/N_c$, which implies that there is a Landau pole in the UV and thus this theory lacks an UV completion. The location of the Landau pole is exactly where the dilaton diverges, {\it i.e.}~strictly in the $r \to \infty$ limit.

\section{Thermodynamics}

In this section we will briefly discuss the thermodynamics of the finite temperature back-reacted background given in equation (\ref{ansatz}) and (\ref{solb})-(\ref{solg}). To make connection with earlier works, we will keep our discussion closely analogous to what is discussed in {\it e.g.}~\cite{Bigazzi:2009bk}. The Hawking temperature of the black hole, which is identified with the temperature of the dual field theory, can be obtained by first compactifying the Euclideanized time-direction and then requiring the regularity of the near-horizon metric. The temperature is identified with the inverse period of the compact Euclidean time-direction. Thus we get
\begin{eqnarray}
T = \left. \frac{1}{4\pi} \frac{d}{dr} \left(h^{-1/2}(r) b(r) \right) \right|_{r_H} = \frac{r_H}{\pi L^2} \left( 1 - \frac{\epsilon}{16} \right) \ .
\end{eqnarray}
The entropy of the ten-dimensional black hole is given by
\begin{eqnarray}
s = \frac{2 \pi A_8}{\kappa_{10}^2 V_{\mathbb R^3}} = \frac{27}{32} \pi^2 N_c^2 T^3 \left( 1 + \frac{1}{8} \epsilon \right) + \cO(\epsilon^2) \ ,
\end{eqnarray}
where $A_8$ denotes the eight-dimensional area of the even-horizon transverse to the time and the radial coordinate. In obtaining the above result we have also used the fact that $2\kappa_{10}^2 = (2\pi)^7 \alpha'^4 g_s^2$. Note that, the above result is not completely new in the sense that we could obtain the same by computing the entropy of the background (before taking any back-reaction) and the probe separately and adding them up.

Proceeding along the same direction, we can evaluate the ADM energy of the black hole spacetime using the formula
\begin{eqnarray}
\varepsilon = - \frac{1}{\kappa_{10}^2} \frac{1}{V_{\mathbb R^3}} \sqrt{|G_{tt}|} \int d^8 x \sqrt{{\rm det G_8}} \left( K_T - K_0 \right) \ ,
\end{eqnarray}
where $G_{tt}$ is the time component of the background metric, $G_8$ denotes the metric on the constant time and constant radius hypersurface and finally $K_T$ and $K_0$ denote the extrinsic curvatures of the eight-dimensional hypersurface within the nine-dimensional constant time hypersurface at non-zero and vanishing temperatures respectively. The formula for evaluating the extrinsic curvature is given by
\begin{eqnarray}
K := \frac{1}{\sqrt{{\rm det} G_9}} \partial_{\mu} \left( \sqrt{{\rm det} G_9} \, n^\mu \right) \ , \quad {\rm where} \quad n^\mu = \frac{1}{\sqrt{G_{rr}}} \delta_r^\mu \ ,
\end{eqnarray}
where $G_9$ denotes the metric on the constant time hypersurface.

The subtraction of the zero temperature background should be viewed as a regularization scheme. Note that, in our case, the zero temperature back-reacted solution is likely to be more complicated, so we have not pursued it here. However, we are going to use the above formula for computing the ADM energy and we will simply send $r_H \to 0$ while computing $K_0$. Operationally this will simply get rid of the (quartic) divergence term and thus will amount to effectively regularizing the energy function. Moreover --- even though the back-reaction at zero temperature potentially convolutes the analysis --- far in the UV the probes have qualitatively similar behavior both at zero and at finite temperature. Thus, purely on physical grounds,  it also makes sense to subtract the term corresponding to $K_0$ which is evaluated by sending $r_H \to 0$. With all these  prerequisites, the final result is obtained to be
\begin{eqnarray}
\varepsilon = \frac{81}{64} \pi^2 N_c^2 T^4 \left(1 + \frac{1}{8} \epsilon \right) + \cO(\epsilon^2) \ .
\end{eqnarray}
We observe that the correction due to the presence of the flavor degrees of freedom contributes equally to the entropy and the internal energy. With this result, we can use the standard definitions and rules of thermodynamics to evaluate {\it e.g.}~the specific heat, $c_V := \partial _T \varepsilon$ or the Helmholtz free energy, $F = \varepsilon - s T$.

\section{Probe branes and phase transition}

In this section we will discuss the physics of an additional probe sector, {\it a la mode} \cite{Kuperstein:2008cq}, that can be introduced in the back-reacted background given in (\ref{solb})-(\ref{solg}). Our task here is to analyze how the back-reaction changes the physics in this additional probe sector.

\subsection{ D7 and anti-D7 branes: phase transition}

We will add $N_f'$ additional D7 and anti-D7 branes in this backreacted geometry with $N_f'\ll N_f$ and we will ignore the backreaction of these additional branes. We will study a D7-brane configuration which spans the space-time coordinates $x_\mu$, the radial direction $r$, and the coordinates $\psi,\theta_2,\phi_2$. The transverse directions are given by two-sphere coordinates $\theta_1,\phi_1$. The DBI (Euclidean) action is given by
\begin{equation}
S_{\rm DBI} = \tau_7 \int e^{\Phi(r)} \sqrt{g_8}d^8x \ .
\end{equation}
Here, $g_8$ is the determinant of the induced metric. Assuming $\theta_1=\theta(r)$, $\phi_1=\phi(r)$ we get
\begin{equation} \label{adprobe}
S_{\rm DBI} = \cN_T \int dr e^{f(r)+2 g(r)+\Phi (r)} \sqrt{\frac{1}{6} b(r) e^{2 g(r)} \left(\theta '^2+ \sin ^2 \theta \phi '^2\right)+1}  \ ,
\end{equation}
where $\cN_T = (2 N_f') \tau_7 V_{\mathbb{R}^3} (8\pi^2)/ (9T)$ and $T$ is the background temperature. The factor of $2$ appears because the brane and the anti-brane contribute equally to the thermodynamic free energy. The equations of motion obtained from the above action admit two classes of solutions: Firstly, the parallel-shaped solution
\begin{equation}
\phi(r)= {\rm const} ,  \qquad  \theta(r)= \pi/2 \ ,
\end{equation}
and, secondly,  the U-shaped solution
\begin{align}
& \theta(r)=\frac{\pi}{2} \ , \\
&\phi '(r)=\frac{6 c}{\sqrt {b(r)^2 e^{2 f(r)+8 g(r)+2 \Phi (r)}-6 c^2 b(r) e^{2 g(r)}}} \ , \label{KSsoln}
\end{align}
where $c$ is some constant which can be determined by using the condition that $\phi '(r)$ diverges at some radial point $r=r_0$
\begin{eqnarray} \label{cdet}
c^2 = \frac{1}{6} \left. b e^{2f + 6g + 2 \Phi}\right |_{r=r_0} \ . 
\end{eqnarray}
The asymptotic behavior of the profile function $\phi$ is given by
\begin{eqnarray}
\phi(r) = \frac{\cphi}{2} - \frac{c}{4 r_*^4} + \ldots \ .
\end{eqnarray}
Thus the asymptotic angle separation is the non-normalizable mode (corresponding to a source) and the constant $c$ is the normalizable mode (corresponding to a VEV). The qualitative picture of the possible embedding functions is again similar to the one demonstrated in fig.~\ref{kwprobe}: The parallel-shaped represents a chiral symmetry restored phase and the U-shaped ones represent the symmetry broken phase.

\begin{figure}[h!]
\centering
\includegraphics[scale=0.65]{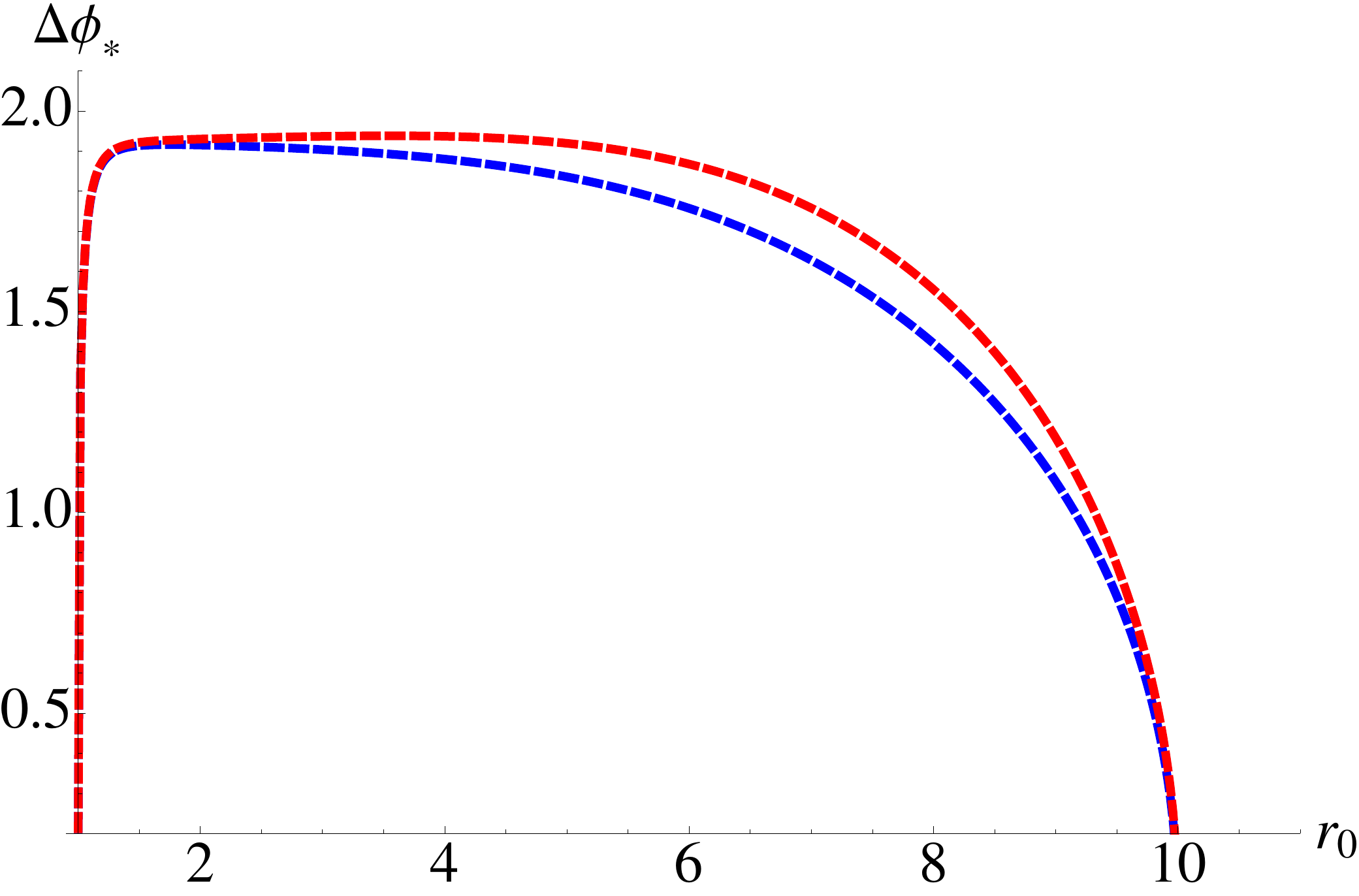}
\caption{\small The dependence of the asymptotic angle separation with the radial variable $r_0$ where the brane--anti-brane pair meets. The blue and red curve corresponds to $\mathfrak{m}^{-2} = 0.06$ and $-0.06$ respectively. Here we have set $r_H=1$, $r_* = 10$ and $\epsilon=0.01$.}
\label{anglesep}
\end{figure}

Before proceeding any further, we can explore how the coupling $\cphi$ depends on the various parameters ({\it e.g.}~$\epsilon$ or $\mathfrak{m}$) in this back-reacted background. A representative plot is shown in fig.~\ref{anglesep}.
A couple of comments are in order: the constant $c$ grows monotonically with $r_0$ and therefore the dependence of $\cphi$ vs $c$ will be qualitatively similar.\footnote{Note that in the thermodynamic ensemble, $\cphi$ and $c$ are canonically conjugate variables and therefore we should be able to conclude about the existence of a phase transition by looking at this plane.} We have chosen to plot $r_0$ since the corresponding values for $c$ are very large numbers.

In \cite{Alam:2012fw} it was demonstrated and subsequently argued that in the original Klebanov-Witten background at finite temperature, the coupling $\cphi$ monotonically increases with increasing $r_0$ and approaches a constant non-zero value where the curve flattens. This in turn means that there can be no phase transition. This is in stark contrast with what we observe here: From fig.~\ref{anglesep} we see that the $\cphi$ curve tends to flatten for $r_0$ considerably larger than the horizon radius; however, then the curve starts decreasing as $r_0$ becomes closer to $r_*$ and eventually goes to zero. This signals that there will be a phase transition once the back-reaction is taken into account. We also observe that for $\mathfrak{m}^{-2} =0.06$ the curve bends faster than for $\mathfrak{m}^{-2} = -0.06$; thus we expect that the critical $\cphi$ --- separating the two phases --- will increase as we go from $\mathfrak{m}^{-2} =0.06$ to $\mathfrak{m}^{-2} = -0.06$. Note that this qualitative difference in behavior is sourced by the fact that we need to use a finite cut-off $r_*$ and therefore it is the presence of the running dilaton in our back-reacted background which changes the physics drastically.

\begin{figure}[h!]
\centering
\includegraphics[scale=0.65]{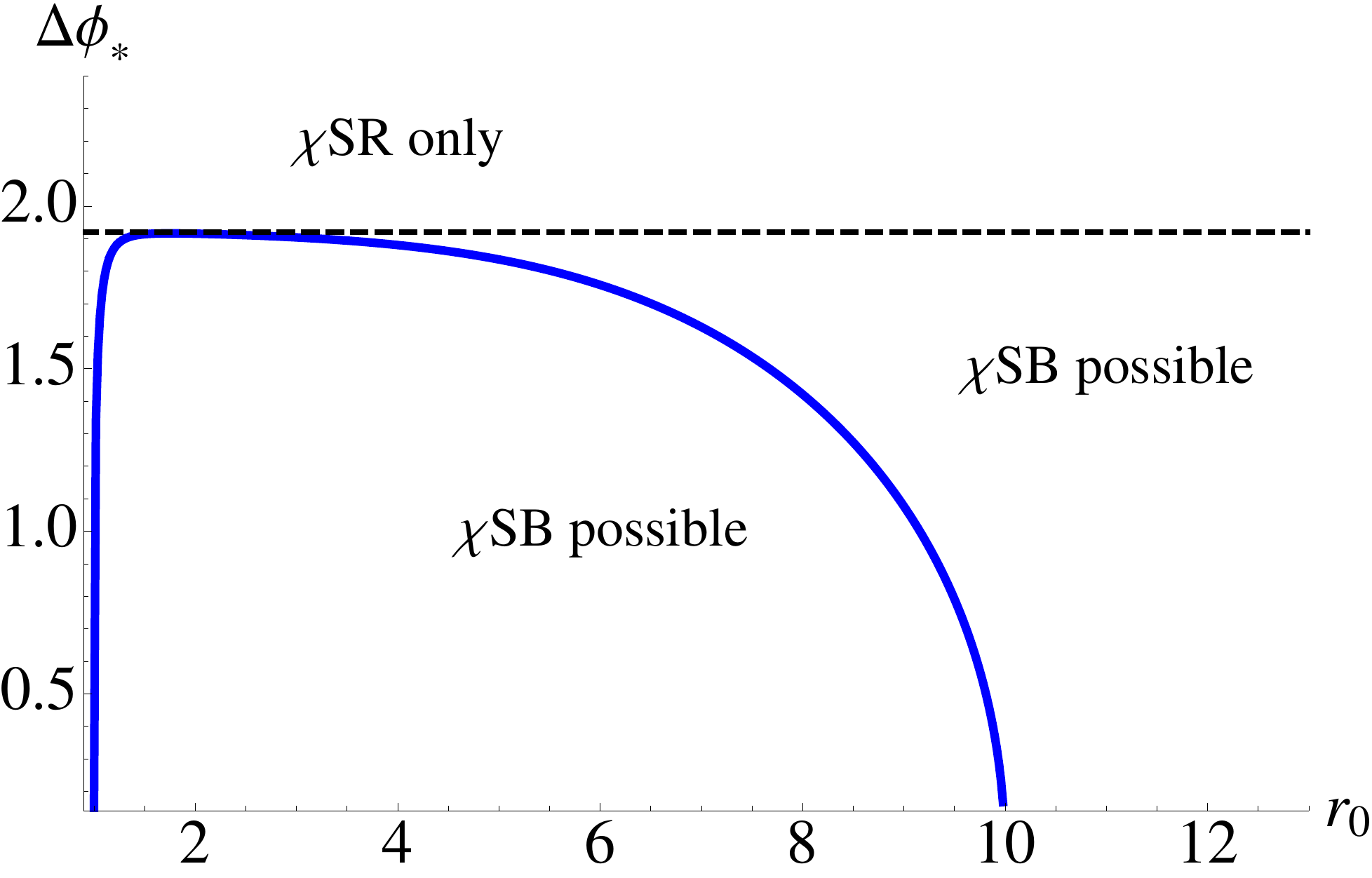}
\caption{\small The dependence of the asymptotic angle separation with the radial variable $r_0$ where the brane--anti-brane pair meets. Here we have set $r_H=1$, $r_* = 10$ and $\epsilon=0.01$ and $\mathfrak{m}^{-2} = 0.06$. The horizontal dashed black line represents the maximum value $\cphi$ beyond which there are no chiral symmetry broken phase.}
\label{delphiphase}
\end{figure}

We can also learn about the existence of possible phases from this curve: this is schematically shown in fig.~\ref{delphiphase}.
Clearly, there exists a maximum $\cphi$ for a given $\mathfrak{m}$ beyond which only chiral symmetry restored phases are available. On the branch where $(\partial \cphi)/(\partial r_0)>0$, the system is thermodynamically unstable. This is because increasing $r_0$ by a small amount will increase the angle separation and will either push the brane--anti-brane pair to infinity or pull them all the way to the horizon. However, note that unlike the case encountered in \cite{Alam:2012fw}, chiral symmetry breaking is possible for the entire range $0 \le \cphi \le \cphi^{\rm max}$, since the $\cphi$ curve bends down all the way to zero.

Now, using the equations of motion we can calculate the on-shell action.\footnote{Recall that in the Euclidean signature, the on-shell action corresponds to the thermodynamic free energy of the corresponding phase.} For the first set of solution, we have
\begin{equation}
S_{||}= \cN_T \int_{r_H}^{\infty} dr e^{f(r)+2 g(r)+\Phi (r)} \ .
\end{equation}
For the U-shaped solutions, we get
\begin{equation}
S_U= \cN_T \int_{1}^{\infty} dr e^{f(r)+2 g(r)+\Phi (r)}\sqrt{\frac{1}{6} b(r) e^{2 g(r)}  \phi '(r)^2+1} \ ,
\end{equation} 
with $\phi '(r)$ given by equation (\ref{KSsoln}). To find out the energetically favored embedding, we need to look at the difference in free energy of these two classes of embeddings:
\begin{equation}
\Delta S = (S_U-S_{||})/\cN_T \ .
\end{equation}
To find out the phase diagram, we will evaluate this quantity numerically.

\begin{figure}[h!]
\centering
\includegraphics[scale=0.65]{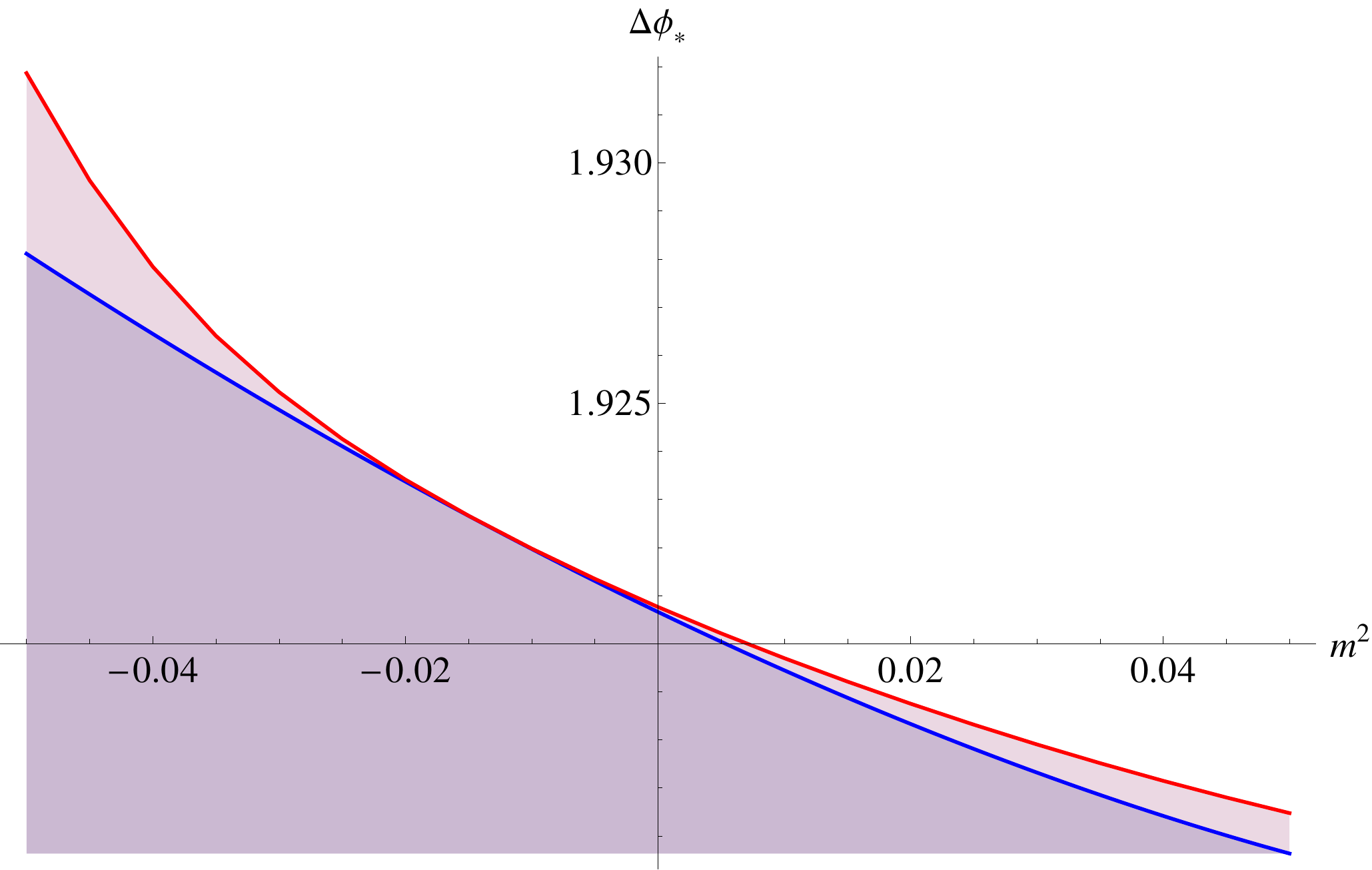}
\caption{\small We have shown the phase diagram in the $\cphi$ vs $m^2$ plane. The blue solid line separates the $\chi$SB $\equiv$ chiral symmetry broken broken phase (below the line) from the $\chi$SR phase (above the line). Above the red solid line only $\chi$SR $\equiv$ chiral symmetry restored phase is available. We have shown these in more details in fig.~\ref{phasediag_details}. We have used $\epsilon = 0.01$, $r_H=1$, $L=1$ and $r_* = 10$.}
\label{phasediag}
\end{figure}

\begin{figure}[h!]
\centering
\includegraphics[scale=0.65]{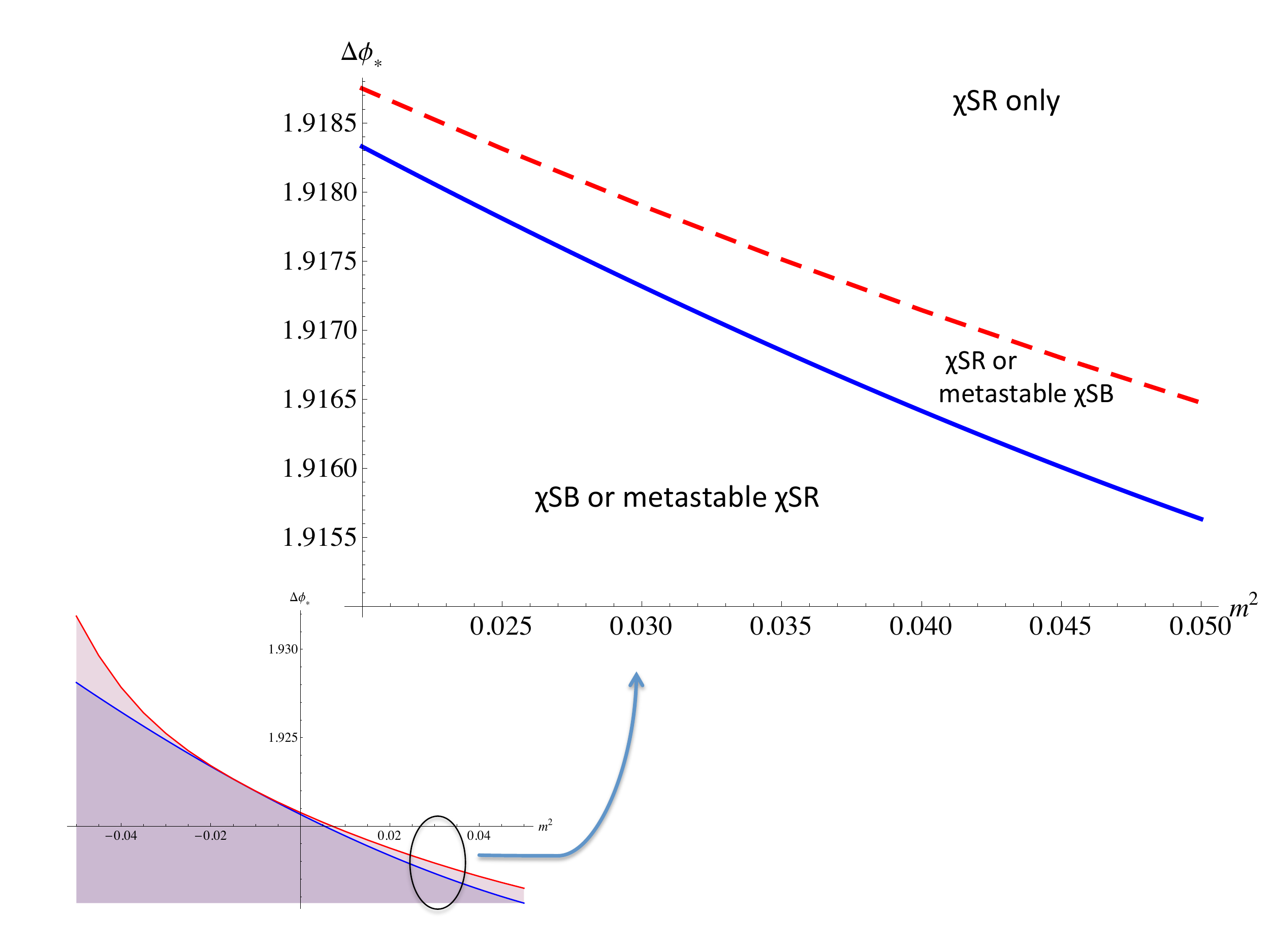}
\caption{\small We have shown the details of a part of the phase diagram presented in fig.~\ref{phasediag}. Above the red dashed line we only have the $\chi$SR phase; between the red dashed  we have both $\chi$SB and $\chi$SR as available phases. The blue solid line separates these phases. We have used $\epsilon=0.01$, $r_H=1$, $L=1$ and $r_*=10$.}
\label{phasediag_details}
\end{figure}

To represent the phase diagram, let us first define the following dimensionless coupling
\begin{eqnarray}
m^2 =  \pi^2 L^4 \frac{T^2}{\mathfrak{m}^2} \left(1 + \frac{\epsilon}{8}\right) \ .
\end{eqnarray}
Thus the coupling $m^2$ measures the relative strength between the background temperature and the dimensionful coupling $\mathfrak{m}$. We will represent the corresponding phase diagram in the $\cphi$ vs $m^2$ plane. The resulting phase diagram is shown in fig.~\ref{phasediag} and fig.~\ref{phasediag_details}.

From fig.~\ref{phasediag} we observe that increasing $m^2$, and therefore increasing temperature with respect to $\mathfrak{m}^2$, favors the chiral symmetry restored phase. This is expected since the introduction of the coupling $\mathfrak{m}^{-2}$ induced a spontaneous breaking of chiral symmetry which would otherwise remain restored at finite temperature. Thus the irrelevant operator that we have introduced {\it via} taking the leading order back-reaction into account has induced this spontaneous symmetry breaking. The detailed phase structure is magnified in fig.~\ref{phasediag_details}. From this we also notice that --- comparing with the corresponding phase diagram that we analyzed in \cite{Alam:2012fw} --- there is no ``runaway or metastable $\chi$SR" phase available when the back-reaction is non-vanishing. Thus the back-reaction ``stabilizes" a part of the phase diagram and makes it physically accessible.

If we are to dig deeper into the origin of the non-trivial phase structure that we obtained here we may be tempted to conclude the following: taking the back-reaction into account has explicitly broken the conformal invariance and therefore we can construct a meaningful quantity, {\it e.g.}~the ratio $T^2/\mathfrak{m}^2$, which can favor one possible phase over the other depending on its strength. This is indeed true; however, it is not the complete story.

Recall that the back-reaction considered in {\it e.g.}~\cite{Bigazzi:2009bk} --- which is different from what we have considered here --- also introduces an irrelevant deformation to the original CFT and introduces a non-trivial scale. Forgetting about the possible interpretation for the moment, we can imagine introducing an additional set of $N_f'$ probe D7 and anti-D7 branes in the finite temperature back-reacted background obtained in \cite{Bigazzi:2009bk}. However, it is straightforward to check that in that case, the so-called $\chi$SR phase always remains favored and a phase transition never happens. This indicates the phase transition we observe here has more to do with the details of the theory, {\it i.e.}~precisely what irrelevant operator we introduce, than just an effect induced by the breaking of the conformal invariance.

\subsection{Thermodynamics of probe D7-branes}

After discussing the physics of the phase transition, we will now discuss the associated thermodynamics in this probe sector.

\subsubsection{Parallel-shaped embedding}

To begin with, we will analyze the parallel-shaped (also sometimes called the black-hole) embedding. In this case, it is possible to obtain analytical results and hence we will discuss this case in detail and then move on to the discussion for the U-shaped profiles. The profile is described by
\begin{equation}
\phi(r)= {\rm const} ,  \qquad  \theta(r)= \pi/2 \ .
\end{equation}
The Helmholtz free energy $F$ is given by $T S_{||}$, where $S_{||}$ is the on-shell Euclidean action for the parallel embeddings:
\begin{equation}
S_{||} = \cN_T \int_{r_H}^{\Lambda}dr e^{f(r)+2g(r)+\Phi(r)} \ .
\end{equation}
The overall constant $\cN_T$ has been defined earlier, {\it e.g.}~below equation (\ref{adprobe}). The on-shell action diverges as the regulator, denoted by $\Lambda$ above, is taken to a large value 
\begin{align}
S_{||} = \cN_T \left[-\frac{2 \Lambda ^6 \left(\mathfrak{m}^{-2} \epsilon \right)}{3}+\frac{1}{192} \Lambda ^4 \left(12 \epsilon  \log \left(\frac{1}{r_*}\right)-12 \epsilon  \log \left(\frac{1}{\Lambda }\right)-7 \epsilon +48\right)\right.\nonumber \\
\left.+\frac{1}{2} \mathfrak{m}^{-2} \Lambda ^2 \epsilon  r_H^4+\frac{1}{192} \left(7 \epsilon  r_H^4-12 \epsilon  r_H^4 \log \left(\frac{r_H}{r_*}\right)-48 r_H^4\right)\right] \ .
\end{align}
Setting the UV-cutoff $\Lambda= r_*$, we get
\begin{align}
S_{||} = \cN_T \left[-\frac{2 r_*^6 \left(\mathfrak{m}^{-2} \epsilon \right)}{3} + \frac{1}{192} r_*^4 \left(-7 \epsilon +48\right) + \frac{1}{2} \mathfrak{m}^{-2} r_*^2 \epsilon  r_H^4 \right. \nonumber\\
\left.  + \frac{1}{192} \left(7 \epsilon  r_H^4-12 \epsilon  r_H^4 \log \left(\frac{r_H}{r_*}\right)-48 r_H^4\right)\right] \ .
\end{align}
Apparently, there are additional $r_*^6$ and $r_*^2$ divergences along with the usual quartic divergence. However, we have to be careful in treating ``diverging" terms which are multiplied by our expansion parameter $\epsilon$. In light of the discussion in (\ref{valid1})-(\ref{valid2}), it is clear that we only have a quartic divergence.\footnote{We will deal with the logarithmic term momentarily.} We can remove the quartic divergence by adding the following counter-term:
\begin{align}
S_{\rm ct} = (2 N'_f \tau_7)  \frac{L}{4} \left(-1+\frac{1}{32}\epsilon - \frac{1}{3}\epsilon r_*^2 \mathfrak{m}^{-2}\right) \int_{r=r_*} d^7 x \sqrt{{\rm det} [\gamma]} \ ,
\end{align}
where $\gamma$ denotes the induced metric on the probe at a constant radial slice. So, finally we have
\begin{align}
S_{||} + S_{\rm ct} = - \cN_T \frac{r_H^4}{8} \left[ \left(\frac{1}{2} \epsilon  \log \left(\frac{r_H}{r_*}\right) + \frac{2}{3} \mathfrak{m}^{-2} r_*^2 \epsilon -\frac{7}{48} \epsilon +1\right)\right] \ .
\end{align}
The background temperature is given by
\begin{equation}
T=\frac{r_H}{\pi L^2}\left(1-\frac{1}{16}\epsilon\right) \ .
\end{equation}
Therefore, using the definition of the 't Hooft coupling $\lambda = 4 \pi g_s N_c$, we get
\begin{align}
S_{||} + S_{\rm ct} = & - s_0 \left( N'_f V_{\mathbb R^3} T^3 \lambda^2\right) \left(1+\frac{1}{4} \epsilon\right) \left[ \frac{1}{2} \epsilon  \log \left(\frac{r_H}{r_*}\right)+\frac{2}{3} \mathfrak{m}^{-2} r_*^2 \epsilon -\frac{7}{48} \epsilon +1\right] \nonumber\\
= & -s_0 \left( N'_f V_{\mathbb R^3} T^3 \lambda^2\right) \left[ 1+\frac{1}{2} \epsilon  \log \left(\frac{r_H}{r_*}\right)+\frac{2}{3} \mathfrak{m}^{-2} r_*^2 \epsilon \right]\left( 1+\frac{5}{48} \epsilon \right) + \cO (\epsilon^2) \ .
\end{align}
Above, we have defined $s_0=\frac{1}{\pi}\left(\frac{3^4}{2^{14}}\right)$. Now we define $\lambda_{\rm eff}(T)$
\begin{equation}
\lambda_{\rm eff}(T) = \lambda \left[ 1+\frac{1}{4} \epsilon  \log \left(\frac{r_H}{r_*}\right)+\frac{1}{3} \mathfrak{m}^{-2} r_*^2 \epsilon \right] + \cO (\epsilon^2) \ .
\end{equation}
This defines for us an effective 't Hooft coupling at finite temperature. Using this effective 't Hooft coupling we get,
\begin{equation}
S_{||} + S_{\rm ct} = - s_0 \left( N'_f V_{\mathbb R^3} T^3 \lambda_{\rm eff}^2\right) \left( 1+\frac{5}{48} \epsilon \right) + \cO(\epsilon^2) \ .
\end{equation}
The free-energy density can be obtained from the on-shell action
\begin{equation}
f_{||} = \frac{F_{||}}{V_{\mathbb R^3}}=-s_0 N'_f T^4 \lambda_{\rm eff}(T)^2\left( 1+\frac{5}{48}\epsilon \right)+ O(\epsilon^2) \ .
\end{equation}
Consequently the entropy density, energy density and pressure are given by
\begin{align}
s_{||} = & -\frac{\partial f_{||}}{\partial T}=4 s_0 N'_f T^3 \lambda_{\rm eff}(T)^2\left( 1+\frac{11}{48}\epsilon \right)+ \cO(\epsilon^2) \ ,\\
e_{||} = & f_{||} + T s_{||} = 3 s_0 N'_f T^4 \lambda_{\rm eff}(T)^2\left( 1+\frac{13}{48}\epsilon \right)+ \cO(\epsilon^2) \ , \\
p_{||} = & - f_{||} = s_0 N'_f T^4 \lambda_{\rm eff}(T)^2\left( 1+\frac{5}{48}\epsilon \right) + \cO(\epsilon^2) \ .
\end{align}
%

\subsubsection{U-shaped embedding}

\begin{figure}[h!]
\centering
\includegraphics[scale=0.70]{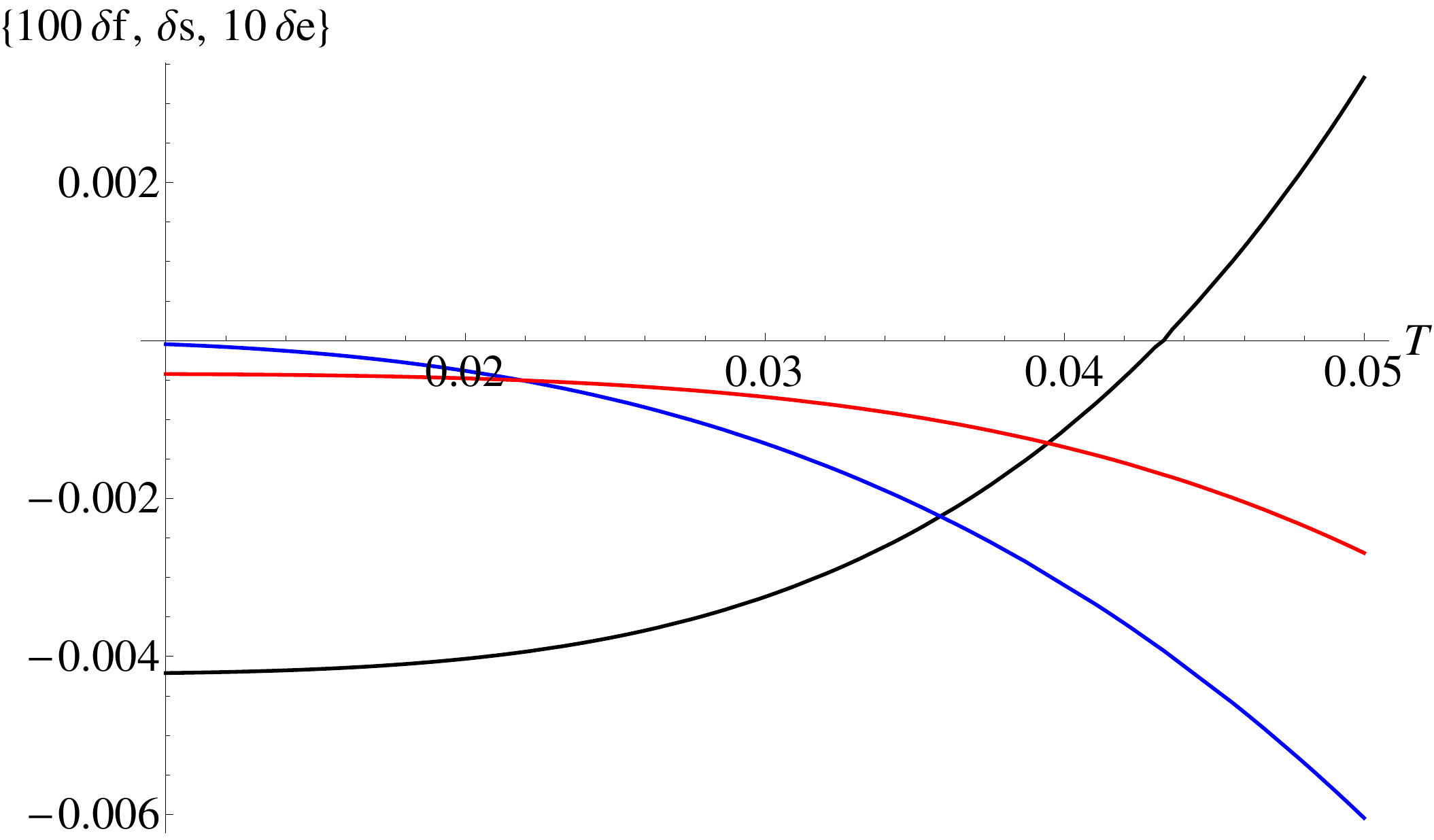}
\caption{Temperature dependence of $100 \delta f(T)$(black), $ \delta s(T)$(blue) and $10 \delta e(T)$(red) for  $\epsilon=0.01$ and $\mathfrak{m}^{-2}=0.01$ and $L=1$. Along the horizontal axis the temperature is measured in units of $r_0$.}
\label{thermo}
\end{figure}

Next, we will discuss the thermodynamics for the U-shaped embedding described by
\begin{align}
& \theta(r)=\frac{\pi}{2} \ , \\
&\phi '(r)=\frac{6 c}{\sqrt{b(r)^2 e^{2 f(r)+8 g(r)+2 \Phi (r)}-6 c^2 b(r) e^{2 g(r)}}} \ , \label{KSsoln2}
\end{align}
where the constant $c$ is determined {\it via} equation (\ref{cdet}). The on-shell action is given by
\begin{equation}
S_U = \cN_T \int_{r_0}^{\infty} dr e^{f(r)+2 g(r)+\Phi (r)}\sqrt{\frac{1}{6} b(r) e^{2 g(r)}  \phi '(r)^2+1} \ .
\end{equation} 
Again all divergences can be removed by adding the counterterm:
\begin{align}
S_{\rm ct} = (2 N'_f) \tau_7 \frac{L}{4} \left(-1+\frac{1}{32}\epsilon-\frac{1}{3}\epsilon r_*^2 \mathfrak{m}^{-2}\right) \int_{r=r_*} d^7 x \sqrt{{\rm det}[\gamma]} \ .
\end{align}
Now all the thermodynamic quantities can be computed:
\begin{align}
f_U = & f_{||} + \left(\frac{3^4}{\pi^3 2^{11}}\lambda^2 N'_f\right)\delta f(T) \ , \\
s_U = & s_{||} + \left(\frac{3^4}{\pi^3 2^{11}}\lambda^2 N'_f\right) \delta s(T) \ , \\
e_U = & e_{||} + \left(\frac{3^4}{\pi^3 2^{11}}\lambda^2 N'_f\right) \delta e(T) \ .
\end{align}
Here $\delta f(T)$, $\delta s(T)$ and $\delta e(T)$ needs to be evaluated numerically and are shown in figs.~\ref{thermo}. 
From fig.~\ref{thermo} we observe that the free energy difference crosses zero at a certain value of the temperature measured in units of $r_0$, which encodes the existence of the phase transition. We also observe that the entropy is always greater in the chiral symmetry restored phase. This is also expected: in the chiral symmetry restored phase more degrees of freedom are available since they cannot be bound to form a chiral condensate. This fact also gives rise to an increased internal energy for the symmetry restored phase. 
\begin{figure}[h!]
\centering
\includegraphics[scale=0.70]{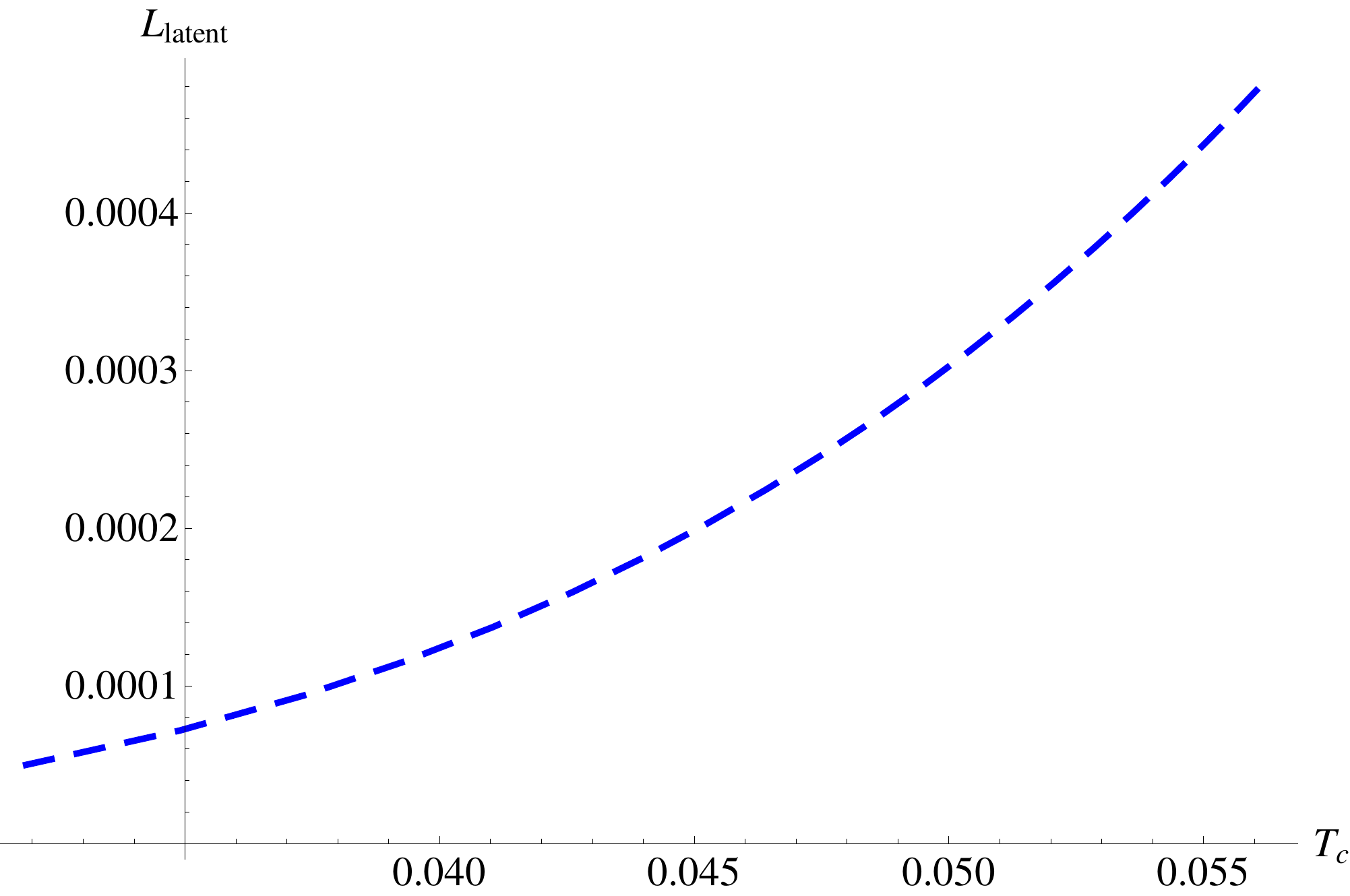}
\caption{Temperature dependence of the latent heat during the phase transition. Here we have used $\epsilon=0.01$, $\mathfrak{m}^{-2}=0.01$, $L=1$ and we have also set $r_0=1$.}
\label{latent}
\end{figure}

A first order phase transition is accompanied by a jump in the entropy density at the critical temperature that results in a non-zero latent heat. This latent heat is simply given by
\begin{eqnarray}
L_{\rm latent} = T_c \, \delta s(T_c) \ .
\end{eqnarray}
On dimensional grounds, the overall scaling of the latent heat should behave as $T_c^3$. However, here we have other scales in the system and thus the most general behavior the latent heat can follow will be of the following form
\begin{eqnarray}
L_{\rm latent} = T_c^3 \, F\left( \frac{T_c^2}{\mathfrak{m}^2} , \frac{T_c}{\Lambda}\right) \ , 
\end{eqnarray}
where $F$ is some unknown function and $\Lambda$ is the UV cut-off of our theory. In fig.~\ref{latent} we have shown a representative behavior of the latent heat with the critical temperature.

\section{Some comments about the stability of the additional probes}

Since we are dealing with non-supersymmetric probes to begin with, we would like to understand whether after taking into account the effect of the back-reaction, this system remains stable or not. Ideally speaking, one needs to study the fluctuations of the full background to make any concrete claims about the stability. We will, however, do the following: we assume that any possible instability in this system will be sourced by the non-supersymmetric probes and by analyzing the stability of the additional set of probe branes in the back-reacted background, we should be able to see the effect of this instability, if there is any. Here we will perform suggestive calculations to shed light on the issue of stability of the additional probe branes.

\subsection{Computing the force on the branes}

One way to ensure stability of probe branes is to simply embed them in a supersymmetric way, which is not the case we currently have. One easy way to see {\it e.g.}~a supersymmetric ``extra" set of probe Dp-branes is stable in the background of a large number of branes of the same dimension is to compute the probe action assuming that the probe brane is ``moving slowly" in the transverse directions. In this case, a cancellation occurs between the Dirac-Born-Infeld piece and the Chern-Simons piece and the resulting Lagrangian takes the form of that of a free particle at the quadratic order in velocity.

We could ask the same question in our non-supersymmetric set-up as well. What we will try to explore here is what happens if we give this additional set of probes some velocity along its transverse directions. In our back-reacted background, we do have a ``large" number of D7-branes, but we do not have any non-trivial RR-potential sourced by them since there is an equal number of D7 and anti-D7 branes uniformly smeared. Thus the additional probe D7-brane will not have any contribution from the Chern-Simons term and it's action is entirely given by the DBI piece. Thus, even in the small velocity approximation, the effective Lagrangian will not be that of a free particle -- but will be of a particle moving in a non-trivial potential. Our goal here is to determine this potential.

To make this more precise let us consider, in the back-reacted background, an additional probe D7-brane whose embedding is given by
\begin{eqnarray} \label{moveprobe}
\theta = \theta(r) + v_\theta t  \ , \quad \phi = \phi(r) + v_\phi t \ .
\end{eqnarray}
Thus we are considering a very general embedding where both $\theta$ and $\phi$ can be a function of the radial coordinate. We will try to extract the Lagrangian in the limit when the velocities are small. Note that, in the background there is an U(1) isometry along the direction $\phi$. Thus without any loss of generality we can set $v_\phi = 0$.

Now, the Lagrangian that results from the above ansatz is complicated looking. However, one can check the following: in the limit of small velocities, the leading order term (which is independent of the velocities) determine the classical profile of the probe and not surprisingly a simple solution for the classical probe is given by
\begin{eqnarray}
\theta_{(0)}(r) = \pi/2 \ , \quad \phi_{(0)}(r) = {\rm const} \ .
\end{eqnarray}
Here the subscript stands for the classical solution. To have a systematic expansion, in the small velocity limit, the profiles will deform and we denote this by
\begin{eqnarray}
\theta = \pi/2 + \alpha \left( \theta_{(1)} (r) + v_\theta t \right) \ , \quad \phi = {\rm const} + \alpha \phi_{(1)}(r) \ ,
\end{eqnarray}
where $\alpha$ is a small parameter. Now it can be explicitly checked that the DBI Lagrangian takes the following form
\begin{eqnarray}
\cL = \cL_{(0)}(r)  + \cL_{(1)}\left(v_\theta, \theta_{(1)}', \phi_{(1)}', r \right) \alpha^2 + \cO(\alpha^4) \ .
\end{eqnarray}
Thus the equations of motion obtained from the Lagrangian at the order $\alpha^2$ gives the solution
\begin{eqnarray}
\theta_{(1)}' = 0 \ , \quad \phi_{(1)}' = 0 \ .
\end{eqnarray}
Furthermore, there is no non-trivial potential (which depends on the direction $\theta$) and thus the velocity $v_\theta$ remains a constant of motion. Thus, a small velocity along either the $\phi$ or the $\theta$ direction does not get accelerated suggesting that the embeddings are possibly stable.

It turns out that one can generalize this argument to a full non-linear level. With the ansatz in (\ref{moveprobe}), the DBI Lagrangian is given by
\begin{eqnarray} \label{dbinl}
\cL & = & \frac{1}{108} e^{f+2g+\Phi} \left[ 6 b e^{2g} \theta'^2 + \frac{1}{b} \left( 6b - e^{2g} h v_\theta^2 \right) \left( 6 + be^{2g} \sin^2\theta \phi'^2 \right) \right]^{1/2} \nonumber\\
& = & \cL(\theta, \theta', \dot{\theta}, \phi') \ , \quad \dot{\theta} \equiv v_\theta \ .
\end{eqnarray}
The equation of motion will now imply
\begin{eqnarray}
\frac{\partial \cL}{\partial \phi'} = c = {\rm const} \ .
\end{eqnarray}
Note that unlike the small $v_\theta$ case, this constant $c\not = 0$ even for the probes that go all the way to the horizon.\footnote{Note that the solution corresponding to $c=0$ will run into trouble in the deep IR. It represents a brane that is moving with some velocity along the $\theta$-direction at the asymptotic boundary. This velocity will be red-shifted as we go closer to the horizon and eventually exceed the speed of light. To prevent this, the brane will bend and develop a non-trivial profile {\it e.g.}~along the $\phi$ direction. Thus $\phi(r)$ will not remain a constant all the way. Such an effect has been discussed in {\it e.g.}~\cite{Albash:2006bs}. However, in the small $v_\theta$ expansion this bending effect can be neglected and we can approximate the brane having a straight shape.}

The other equation of motion will be given by
\begin{eqnarray}
\frac{d}{dt} \left(\frac{\partial \cL}{\partial \dot{\theta}}\right) + \frac{d}{dr} \left(\frac{\partial \cL}{\partial \theta'}\right) - \frac{\partial \cL}{\partial \theta} = 0 \ .
\end{eqnarray}
Using the explicit form of the Lagrangian in (\ref{dbinl}), it can be shown that we still have the following solution
\begin{eqnarray}
\theta(r) = \pi/2 \ , \quad \frac{d}{dt} v_\theta = 0 \ .
\end{eqnarray}
Hence the static probes, when given an initial velocity along the angular directions, will not accelerate. This also suggests that the probe sector is indeed stable.

\subsection{Asymptotic behavior of the profile functions}

So far we have focussed on the solution $\theta = \pi/2$ and $\phi = {\rm const}$. But we can also look for non-constant solutions and from their asymptotic fall-off behavior we can read off the dimension of the corresponding operator. Typically in AdS/CFT, an instability (due to non-unitarity) of the dual gauge theory is reflected in the fact that the dimension of this operator becomes complex-valued.

Although the back-reacted solution does not asymptotically approach AdS$_5\times T^{1,1}$, the AdS$_5$ part is still present. Thus we can use the above intuition to explore whether a non-constant solution for either $\theta$ or $\phi$ has a problematic UV behavior.

The equation of motion for $\phi$ is given by
\begin{eqnarray}
\frac{e^{f+4g+\Phi} b \sin^2\theta \phi'}{\left[1 + (1/6) b e^{2g} \left(\theta'^2 + \sin^2\theta \phi'^2 \right) \right]^{1/2}} = c_\phi \ .
\end{eqnarray}
The equation of motion for $\theta$ is given by
\begin{eqnarray}
\frac{d}{dr} \left[ \frac{e^{f+4g+\Phi} b\theta' } {\left[1 + (1/6) b e^{2g} \left(\theta'^2 + \sin^2\theta \phi'^2 \right) \right]^{1/2}} \right] - \frac{b e^{f+4g+\Phi} \sin(2\theta) \phi'^2}{\left[1 + (1/6) b e^{2g} \left(\theta'^2 + \sin^2\theta \phi'^2 \right) \right]^{1/2}} =  0 \ .
\end{eqnarray}

Now let us first check the case: $\theta = \pi/2$ and $\phi = {\rm const} + \phi(r)$. Assuming that $\phi' \to 0$ fast enough as $r \to \infty$ such that we can ignore the $\phi'$-dependent term under the square root in the denominator, we get
\begin{eqnarray}
\phi' = \frac{8 c_\phi}{r^5 \left( 8 - \epsilon + 2 \epsilon \log r \right)}  \ .
\end{eqnarray}
Using a small $\epsilon$ expansion, the asymptotic behavior can now be obtained to be
\begin{eqnarray}
\phi = {\rm const} + c_\phi \frac{4 \epsilon \log r - 16 - \epsilon}{64 r^4} + \ldots \ .
\end{eqnarray}
Thus the asymptotic behavior picks up a logarithmic term, but it does not have any monomial in $r$ whose exponent is complex-valued.

The other simplifying case we can explore is to consider $\theta = \pi/2 + \theta(r)$ and $\phi = {\rm const}$. In this case also one can show
\begin{eqnarray}
\theta = \pi/2 + c_\theta \frac{4 \epsilon \log r - 16 - \epsilon}{64 r^4} + \ldots \ .
\end{eqnarray}
Actually we can simultaneously consider $\theta = \pi/2 + \theta(r)$ and $\phi = {\rm const} + \phi(r)$ and using the equations of motion it is straightforward to argue that their UV behavior are exactly the same as we have written here. The above argument is fairly suggestive that the probe should be stable.

\subsection{Studying the fluctuations}

Let us now study the fluctuation equations obtained in (\ref{scalarfluc1}) and (\ref{scalarfluc2})
\begin{eqnarray}
&& \partial_a \left[e^{\Phi} \sqrt{-E^0} g_0^{ab} G_{\theta_1\theta_1}^0 \partial_b \delta \theta_1\right] + e^{\Phi} \sqrt{-E^0} \partial_\psi \delta\phi_1 = 0 \ , \\
&& \partial_a \left[e^{\Phi} \sqrt{-E^0} g_0^{ab} G_{\phi_1\phi_1}^0 \partial_b \delta \phi_1\right] - e^{\Phi} \sqrt{-E^0} \partial_\psi \delta\theta_1 = 0 \ .
\end{eqnarray}
Here $G^0$ denotes the background metric evaluated at the classical profile of the probe, $g_0$ denotes the induced metric of the classical profile, $E^0$ denotes the on-shell determinant of the DBI-part of the action and the indices $a, b$ run over the worldvolume coordinates of the probe brane.

It is straightforward to check that for the given background, both the fluctuations obey the same equations of motion. We will henceforth denote this generic fluctuation mode by $\delta X$. It is difficult to solve the above two equations in complete generality. It is also not possible to have separation of variables in the most general case. Thus we need to make some simplifying assumptions to make any progress. \\

\noindent Case 1. We begin by assuming
\begin{eqnarray}
\partial_\psi \delta X = 0 \  ,
\end{eqnarray}
where $\delta X$ represents any of the fluctuation modes. Note that this simplification should not compromise the generality of our discussion. Even after introducing the additional probe sector, the symmetry along the $\psi$-direction is not affected and thus we do not expect any instability to arise along that direction.

Furthermore, we assume that the fluctuations are independent of the Minkowski directions. With this simplification we can separate the variables\footnote{If we assume a plane-wave type ansatz for oscillations along the Minkowski directions, then we can also separate the variables.} and the resulting equations are obtained to be
\begin{eqnarray}
&& \partial_r \left[ e^{f+4g+\Phi} b \partial_r Y(r) \right] - 6 \ell (\ell+1)e^{f+2g+\Phi} Y(r) = 0 \ , \\
&& \nabla^2_{S^2} Z + \ell (\ell+1) Z = 0 \ .
\end{eqnarray}
Here we have used $\delta X = Y(r) Z(\theta_2, \phi_2)$.

We can recast the radial equation of motion in the following form
\begin{eqnarray} \label{sch1}
\sqrt{\frac{a_1}{a_3}} \partial_r \left( \sqrt{\frac{a_1}{a_3}} \partial_r \kappa (r) \right) - 6 \ell(\ell+1) \kappa(r) - V(r) \kappa(r) = 0 \ ,
\end{eqnarray}
where 
\begin{eqnarray}
Y(r) = \sigma(r) \kappa(r) \ , 
\end{eqnarray}
with
\begin{eqnarray}
\frac{\sigma'}{\sigma} = - \frac{1}{2} \left[ \frac{a_2}{a_1} + \frac{1}{2} \frac{a_1}{a_3} \partial_r \left(\frac{a_3}{a_1}\right) \right] \ .
\end{eqnarray}
And also,
\begin{eqnarray}
&& a_1(r) =  e^{f+4g+\Phi} b \ , \quad a_2(r) = \partial_r \left( e^{f+4g+\Phi} b \right) \ , \quad a_3(r) =  e^{f+2g+\Phi} \ , \\
&& A(r) = \frac{a_2}{a_1} \ , \quad B(r) = \left(\frac{a_3}{a_1}\right)^{1/2} \ , \\
&& V(r) = - \frac{1}{B^2} \left[\frac{1}{4} \left(A + \frac{B'}{B}\right)^2 - \frac{1}{2} \partial_r \left(A + \frac{B'}{B}\right) - \frac{A}{2} \left(A + \frac{B'}{B}\right)\right] \ .
\end{eqnarray}
Here $' \equiv \partial_r$ everywhere.

\begin{figure}[!ht]
\begin{center}
\subfigure[] {\includegraphics[angle=0,
width=0.45\textwidth]{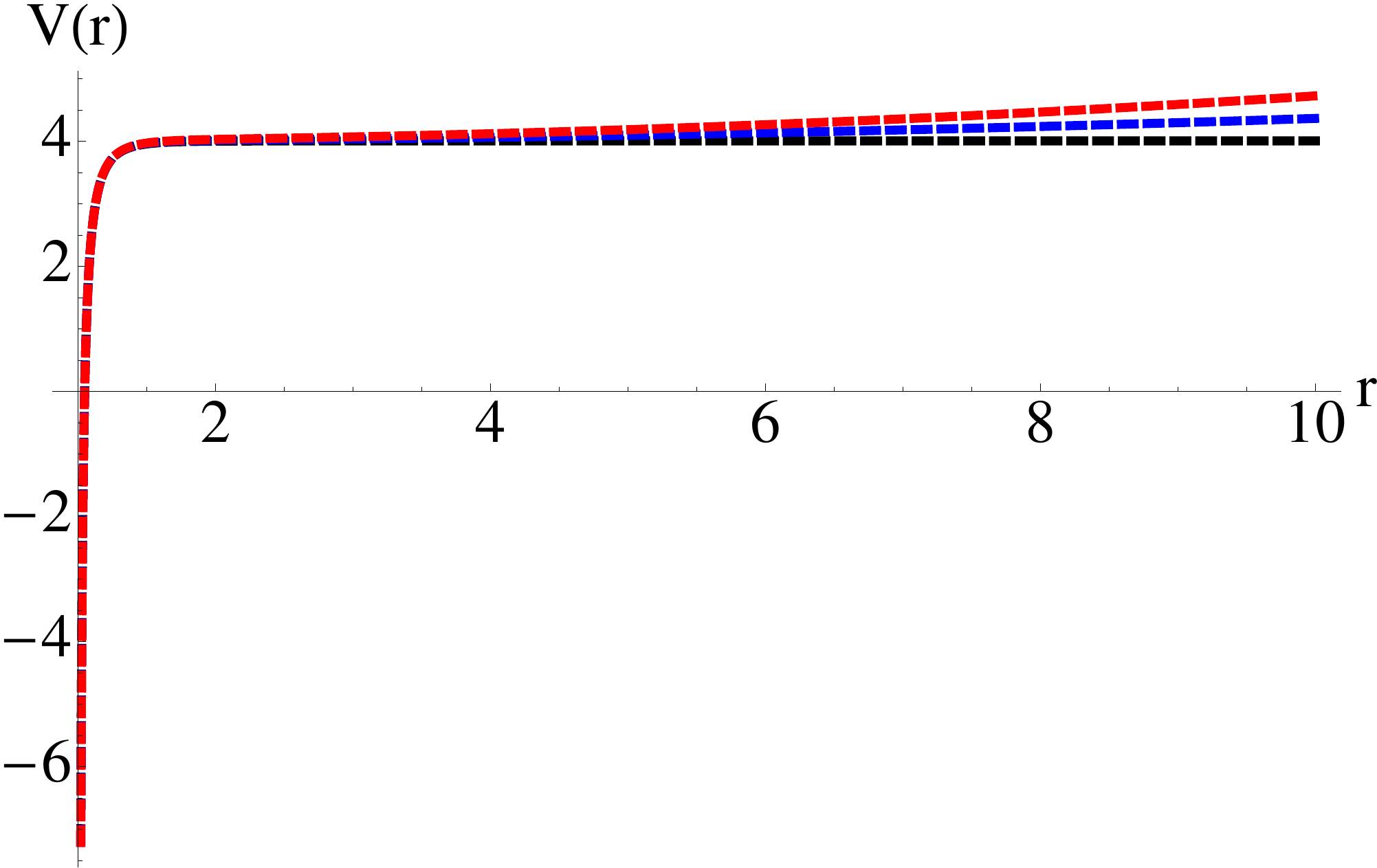} } 
\subfigure[] {\includegraphics[angle=0,
width=0.45\textwidth]{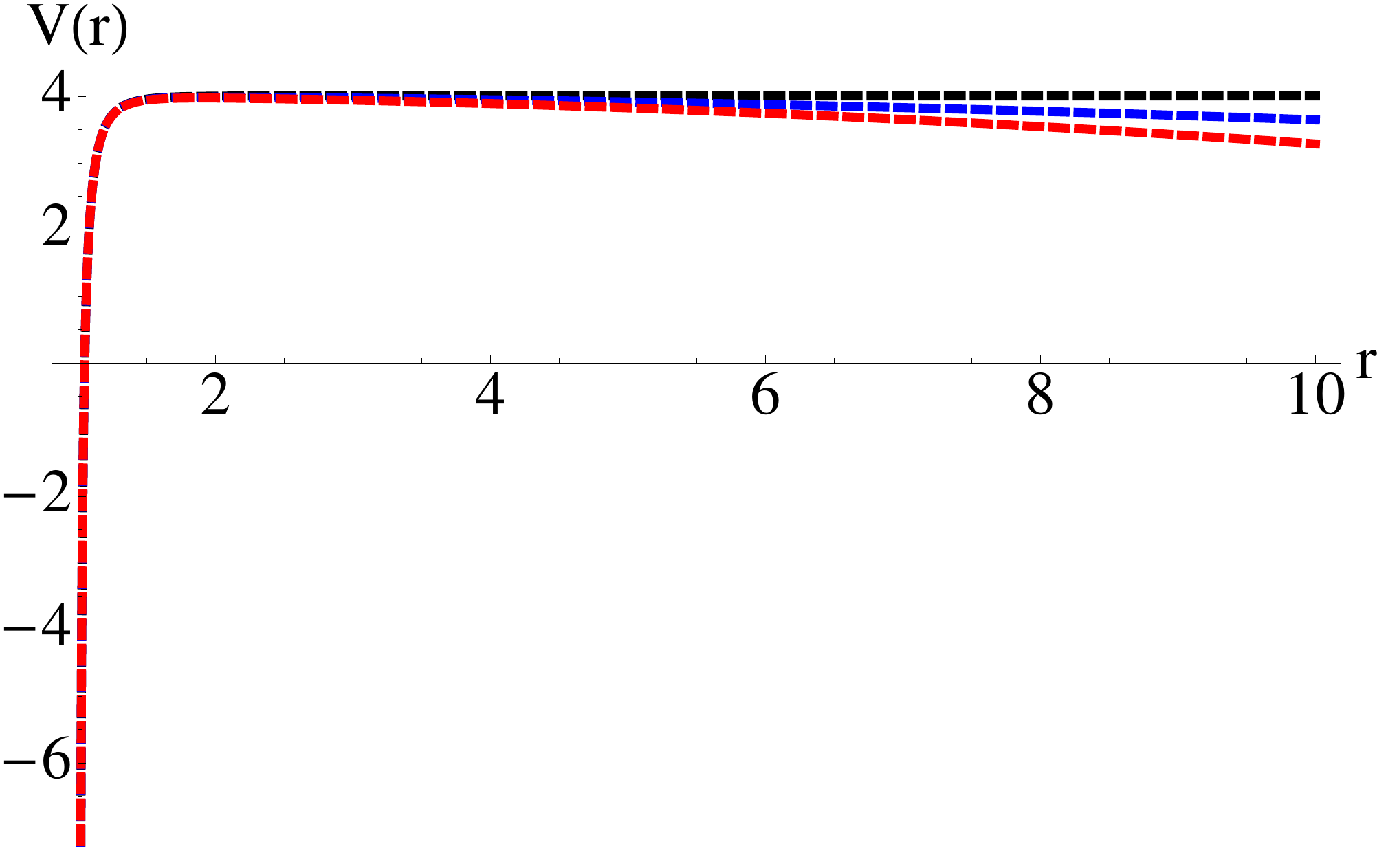} } 
\caption{\small We fix $\epsilon=0.01$. Left panel: The behavior of the potential for various positive values of $\mathfrak{m}^{-2}$. The black, blue and red curves correspond to $\mathfrak{m}^{-2}=0, 0.03, 0.06$ respectively. Right panel: The behavior of the potential for various negative values of $\mathfrak{m}^{-2}$. The black, blue and red curves correspond to $\mathfrak{m}^{-2}=0, -0.03, -0.06$ respectively.}
\label{fluc1}
\end{center}
\end{figure}

The advantage of expressing the fluctuation equation in the form given in (\ref{sch1}) is that it takes the form of a Schr\"odinger equation with a given potential denoted by $V(r)$. By looking at the behavior of this effective potential one can draw conclusions about the nature of the fluctuations. Now we can look at how the effective potential behaves as we introduce the parameter $\epsilon$ and the constant $\mathfrak{m}$.\footnote{Note that in this case these two parameters appear in a single combination $(\epsilon \mathfrak{m}^{-2} r_H^2)$.} This is shown in fig.~\ref{fluc1}.
From fig.~\ref{fluc1} it is clear that for $\mathfrak{m}^{-2}<0$ the effective potential tends to develop a runaway behavior for large $r$. For $\mathfrak{m}^{-2}>0$, however, the effective potential does not develop any possibly pathological behavior as compared to the case of vanishing back-reaction. Actually it can be shown analytically that for large value of $r$, the effective potential behaves as
\begin{eqnarray}
V \to  \left(4 + \frac{\epsilon}{3} \right) + 12 \mathfrak{m}^{-2} \epsilon r_*^2 + \cO(r^{-2}) \ ,
\end{eqnarray}
where $r_*$ is the UV cut-off. Now, recall that the condition that our perturbative solution is valid sets a constraint that $\epsilon |\mathfrak{m}^{-2} r_*^2|$ is parametrically much smaller compared to an order one number. This means that the effective potential should not become negative even for $\mathfrak{m}^{-2}<0$. It is an interesting question how this effective potential might look like when $\epsilon$ is not a small parameter. To answer this question, one needs to find the full solution of the equations of motion and this most likely calls for a numerical endeavor. Currently, we do not have enough information to comment on this situation. On the other hand, $\mathfrak{m}^{-2} > 0$ clearly should not have any instability. \\

\noindent Case 2: Let us now assume that the fluctuations are independent of the angular coordinates. Since there is an SO(3) symmetry, without any loss of generality, we can assume the fluctuations to depend on only one of the spatial directions.

In this case, the equation of motion is given by
\begin{eqnarray}
\partial_r \left( e^{f+4g+\Phi} b \partial_r \delta X\right) - \frac{h}{b} e^{f+4g+\Phi} \partial_t^2 \delta X + h e^{f+4g+\Phi} \partial_x^2 \delta X = 0 \ .
\end{eqnarray}
Assuming a plane-wave type ansatz for the oscillations along the Minkowski directions the equation of motion can be obtained to be
\begin{eqnarray}
&& \partial_r \left( e^{f+4g+\Phi} b \partial_r Y(r) \right) + \omega^2 \frac{h}{b} e^{f+4g+\Phi} Y(r)  - k^2 h e^{f+4g+\Phi} Y(r) = 0 \ , \\
&& {\rm where} \quad \delta X = Y(r) e^{-i\omega t + i k x} \ .
\end{eqnarray}
%

\subsubsection{Case of $\omega = 0$}

We can again recast the radial equation in an effective Schr\"odinger equation with an effective potential. The process is very similar to the one outlined before, except the definition of the function $a_3$ will be different in this case, {\it i.e.}
\begin{eqnarray}
a_3(r) = h e^{f+4g+\Phi} \ .
\end{eqnarray}
The equation of motion now takes the form
\begin{eqnarray}
\sqrt{\frac{a_1}{a_3}} \partial_r \left( \sqrt{\frac{a_1}{a_3}} \partial_r \kappa (r) \right) -  k^2 \kappa(r) - V(r) \kappa(r) = 0 \ ,
\end{eqnarray}
where
\begin{eqnarray}
Y(r) = \sigma(r) \kappa(r) \ .
\end{eqnarray}
In this case the effective potential is given by
\begin{eqnarray}
V(r) = \frac{\epsilon  \left(r_h^4+r^4\right){}^2}{32 r^2 L^4 \left(r^4-r_h^4\right)}-\frac{18 r^4 r_h^4+r_h^8-15 r^8}{4 r^2 L^4
   \left(r^4-r_h^4\right)} \ .
\end{eqnarray}
It is straightforward to check that this potential does not develop any qualitatively new behavior as $\epsilon$ is dialed up. Of course, this is completely insensitive to the choice of the constant $\mathfrak{m}^{-2}$.

\subsubsection{Case of $k=0$}

In this case, the function $a_3(r)$ is given by
\begin{eqnarray}
a_3(r) = \frac{h}{b} e^{f+4g+\Phi} \ .
\end{eqnarray}
Here the equation of motion takes the form
\begin{eqnarray}
\sqrt{\frac{a_1}{a_3}} \partial_r \left( \sqrt{\frac{a_1}{a_3}} \partial_r \kappa (r) \right) + \omega^2 \kappa(r) - V(r) \kappa(r) = 0 \ ,
\end{eqnarray}
with the potential given by
\begin{eqnarray}
V(r) = \frac{3 \left(r^4-r_h^4\right) \left(3 r_h^4+5 r^4\right)}{4 r^6 L^4}+\frac{\epsilon  \left(-10 r^4 r_h^4+9 r_h^8+r^8\right)}{32
   r^6 L^4} \ .
\end{eqnarray}
Again, it can again be checked explicitly that this potential does not develop any new qualitative feature as we increase $\epsilon$. Thus there is no possibility of an instability within this analysis either.

\section{Conclusions and future directions}

In this article we have studied the effects of back-reaction of flavor D7-/anti-D7-branes in the so-called Kuperstein-Sonnenschein model at finite temperature. We have treated the back-reaction perturbatively in which case it turns out to be possible to obtain an analytical solution for the background geometry. This can be regarded to be the first step towards a more involved computation that will address the back-reaction non-perturbatively. Such an endeavor most likely calls for a numerical exploration where one needs to find series solutions for various unknown functions near the event-horizon and the asymptotic boundary and eventually construct a numerical solution that successfully interpolates between these two limits. Such series solutions, especially near the asymptotic boundary, are rather non-trivial and take a general form discussed in \cite{Aharony:2007vg}. It will be an interesting exercise to explore the numerical solution to learn more about the physics when $N_f/N_c$ is of arbitrary strength. This will shed light on the effect of the back-reaction for arbitrary values of $N_f/N_c$ on the phase diagram and the question of stability beyond the probe limit.

Let us offer some comments regarding the stability issue. Before taking the back-reaction the na\"{i}ve picture for the construction is rather simple: we have localized probe branes and anti-branes located at two points in the $\phi$-direction. They are equivalent to having $N_f$ and $-N_f$ charges at two points on a circle. A simple electrostatic intuition tells us these charges should attract each other and any little perturbation is likely to drive them towards each other and destabilize the system. However, our analysis does not confirm  this intuition. Note that, contrary to what has been observed in \cite{Burrington:2007qd}, we do not see any non-trivial cancellation between the DBI and the Chern-Simons terms in the quadratic fluctuation for the additional probe sector. Thus the stability that we observe is likely rooted in the fact that we have treated the back-reaction perturbatively. The mass spectrum of the system at finite temperature must be set by the scale of the background temperature. A deformation, brought about by a small perturbation, is unlikely to drive this mass spectrum to become tachyonic. This insinuates a limitation of our approach.

On the other hand, the lack of a non-trivial cancellation as observed in \cite{Burrington:2007qd} can be attributed to the use of the smearing technique which does not generate any non-trivial axion field to which the additional probes can couple to. Going beyond the smearing technique, one feasible scenario is that such a cancellation eventually guarantees stability for the system. Thus it is tempting to conjecture that in this system, in which supersymmetry has absolutely no role to play, various forces do perfectly balance each other to guarantee the desired stability. It will be interesting to understand whether such statements can be made more rigorous.

As much as the above-mentioned directions of future work are promising to be very fruitful, there is more interesting physics to explore within the current setup. The additional probe sector introduced in the back-reacted background can teach us interesting lessons about this model. In \cite{Alam:2012fw}, we have explored the phase structure of the Kuperstein-Sonnenschein model at finite temperature and in the presence of an external electro-magnetic field. We can analyze similar physical questions, now including the effect of the back-reaction perturbatively. This will elucidate how the phase diagrams depend on the various parameters beyond probe limit computations. Another direction to pursue would be to include the effect of a chemical potential in this model. In the presence of such external parameters, we expect to recover a similar physical picture as reported in {\it e.g.}~\cite{Albash:2007bk, Albash:2007bq, Erdmenger:2007bn, Bergman:2008sg, Johnson:2008vna, Johnson:2009ev}. Since the devoted reader has already experienced a fair share of agony and ecstasy, we will report on these in our forthcoming works \cite{forthcoming}.

Moreover, one could consider studying the back-reaction including the various gauge field excitations on the probes. Such gauge field excitations again correspond to a chemical potential or a constant electromagnetic field. Work along these directions has been performed in other models in {\it e.g.}~\cite{Bigazzi:2011it, Filev:2011mt, Ammon:2012qs}. It will be interesting to investigate and identify the robust universal features within our model as well.

Note that the physics of chiral symmetry breaking can --- perhaps more interestingly --- also be realized by placing probe brane--anti-branes in the Klebanov-Strassler background     \cite{Dymarsky:2009cm}. However, the finite temperature version of the background is known only numerically in \cite{Aharony:2007vg} or approximately for large temperatures\cite{Gubser:2001ri}. The problem of considering back-reaction in this case will be technically more involved, but perhaps more rewarding from a physics point of view. Interestingly, in that model, the antipodal probe solution, simply described by $\phi = {\rm const}$, exists even at zero temperature. Thus analyzing the back-reaction for such profiles may be a tractable and interesting problem to consider.

Our work crucially depends on the smearing technique and it really captures the physics in an approximation --- akin to the ``s-wave" approximation --- of the actual situation. Indeed we really have localized sources and not smeared sources. Although we realize interesting physics within this approximation, it is not evident how robust these features actually are beyond this framework, which is a technically more challenging problem because one has to consider localized sources and thus one will have to work with coupled PDEs instead of the coupled ordinary differential equations (ODE) that we have encountered herein. Nonetheless, it will be an interesting problem to consider in the future.\footnote{For some work from a slightly different perspective, where such localized sources have been considered, see {\it e.g.}~\cite{Grana:2001xn, Bertolini:2001qa, Burrington:2004id, Kirsch:2005uy}.}

Let us conclude by saying that although the problem of understanding the strongly coupled physics of the quark-gluon plasma remains a daunting task, we are slowly making progress in uncovering some key universal features of the strong coupling regime using these model computations.

\section*{Acknowledgments}

We are grateful to Elena Caceres and Carlos Nunez for numerous illuminating discussions and encouragements about this work. We specially thank Javier Tarrio for raising a crucial point regarding the identification of the dimension $8$ operator. MI and AK are grateful to the Kavli Institute for the Physics and Mathematics of the Universe at the University of Tokyo and MIT for hospitality during the final stages of this project. This material is based upon work supported by an IRCSET postdoctoral fellowship (MI), the National Science Foundation under Grant no. PHY-0969020 (AK and SK), a Simons postdoctoral fellowship awarded by the Simons Foundation (AK) and the Texas Cosmology Center (SK), which is supported by the College of Natural Sciences and the Department of Astronomy at the University of Texas at Austin and the McDonald Observatory.

\renewcommand{\theequation}{A.\arabic{equation}}
\setcounter{equation}{0}  
\section*{Appendix A. The smearing form}
\addcontentsline{toc}{section}{Appendix A. The smearing form}

The idea of the smearing is to recover the full symmetry of the background we had before placing any probes. The ${\mathbb Z}_2$-symmetry can be recovered simply by picking another pair of D7/anti-D7 branes with transverse plane described by the $\{\theta_2, \phi_2\}$ directions for each pair of D7/anti-D7 brane with transverse plane described by $\{\theta_1, \phi_1\}$. The induced line element on the radial direction of the probe brane is simply
\begin{eqnarray}
\left[ \frac{L^2}{r^2} + (\theta_{1,2}')^2 + \sin^2 \theta_{1,2} (\phi_{1,2}')^2 \right] dr^2 \ ,
\end{eqnarray}
which is invariant (up to reparametrizations) under the shifts: $\theta_{1,2} \to \theta_{1,2} + \theta_0$ and $\phi_{1,2} \to \phi_{1,2} + \phi_0$ where $\theta_0$ and $\phi_0$ are arbitrary constant numbers. Thus we can initially pick $\theta_{1,2} = \pi/2$ and $\phi_{1,2} = \phi(r)$ to focus on the ``great circle".

We can smear the branes on this great circle by using the shift symmetry in $\phi_{1,2}$; then we can rotate this great circle using the shift symmetry in $\theta_{1,2}$. This process should then cover the entire $S^2$ and thus recover the full symmetry of the background. The density of the branes is then given by
\begin{eqnarray}
\rho_{\theta_{\infty} \phi_{\infty}} = \frac{N_f}{4\pi} \sin\theta_{\infty} \ ,
\end{eqnarray}
where $\theta_{\infty}$ and $\phi_{\infty}$ are the asymptotic values of $\theta_{1,2}$ and $\phi_{1,2}$ respectively.

Now to determine the smearing form we follow \cite{Bigazzi:2008zt}. The most general 2-form invariant under $SU(2)\times SU(2)\times U(1)\times {\mathbb Z}_2$ is given by
\begin{eqnarray}
\Omega & = & a_1(r) \left(e^1\wedge e^2 + \tilde{e}^1 \wedge \tilde{e}^2 \right) + a_2(r) \left( e^1 \wedge \tilde{e}^2 +  \tilde{e}^1 \wedge e^2 \right)  \nonumber\\
& + & a_3(r) \left( e^1  + \tilde{e}^1  \right) \wedge \left(e^3 + \tilde{e}^3 \right) + a_4(r) \left( e^2  + \tilde{e}^2  \right) \wedge \left(e^3 + \tilde{e}^3 \right) + \omega \ ,
\end{eqnarray}
where $a_i$, $i=1, \ldots, 4$ are four hitherto undetermined functions. Here $\omega$ is another 2-form invariant under the same aforementioned symmetry and is independent from the other terms. Now $\omega$ can be fixed in the following by demanding that $\Omega$ is closed, {\it i.e.}~$d\Omega = 0$. After some algebra it is straightforward to show that $\Omega$ takes the following form
\begin{eqnarray}
\Omega = \frac{N_f(r)}{4\pi} \left( \sin\theta_1 d\theta_1 \wedge d\phi_1 + \sin\theta_2 d\theta_2 \wedge d\phi_2 \right) - \frac{N_f'(r)}{4\pi} dr \wedge \left( d\psi + \cos\theta_1 d\phi_1 + \cos\theta_2 d \phi_2 \right) \ ,
\end{eqnarray}
which matches exactly the result of \cite{Bigazzi:2008zt}, including the overall normalization constant.

Now, in the notations of \cite{Bigazzi:2008zt}, the embedding functions of the probe branes are given by
\begin{eqnarray}
f_1 = \theta_{1,2} - \theta_{\infty} \ , \quad f_2 = \phi_{1,2} - \phi(r) \ ,
\end{eqnarray}
where $\theta_{\infty}$ is the asymptotic value of $\theta_{1,2}$ (which is fixed to be $\pi/2$ for a single set of branes). Now $\Omega$ can also be evaluated using the formula \cite{Bigazzi:2008zt}
\begin{eqnarray}
\Omega = \sum_{i=1}^2 \int \rho_{\theta_{\infty} \phi_{\infty}} \delta\left( \theta_i - \theta_{\infty} \right) \delta\left( \phi_i - \phi(r) \right) d\theta_{\infty} d \phi_{\infty} \left( d\theta_i \wedge d \phi_i \right) \ .
\end{eqnarray}
From the above formula it is clear that $N_f'(r) = 0$ (since there is no $r\psi$ -component for example). Thus $N_f(r)$ is simply constant. At non-zero temperature the energetically favored embeddings are the ones which go all the way into the horizon \cite{Alam:2012fw} and therefore $N_f(r) = N_f$ for all values of $r$. On the other hand, at zero temperature we know that the probe banes meet at some radial position given by $r_0$ (we fix this parameter in our problem). So for a given $r_0$, $\Omega = 0$ for $r<r_0$. Thus we conclude
\begin{eqnarray}
N_f(r) & = & N_f \quad \forall r > r_0 \ , \nonumber\\
           & = & 0  \quad \forall r < r_0 \ .
\end{eqnarray}
%

\renewcommand{\theequation}{B.\arabic{equation}}
\setcounter{equation}{0}  
\section*{Appendix B. The equations of motion}
\addcontentsline{toc}{section}{Appendix B. The equations of motion}

To set up the conventions, we will start with the action for type IIB supergravity together with the D7-brane DBI action written in the Einstein frame. In our framework, we have an equal number of D7 and anti-D7 branes uniformly smeared and therefore we do not source any $C_8$ potential (the Hodge dual of the axion field). The action reads
\begin{eqnarray}
S & = & S_{\rm SUGRA} + S_{\rm flavour} \nonumber\\
& = & \frac{1}{2\kappa_{10}^2} \int d^{10}x \, \sqrt{-G_{10}} 
\Big[ R - \frac{1}{2} \partial_M \Phi \partial^M \Phi -\frac{1}{4} |F_5|^2 \Big]  \nonumber\\
& - &  2 \tau_7 \sum^{N_f} \int d^{8}\xi \, e^\Phi \Big[ 
\sqrt{-G_{8}^{(1)}} \Big]  -   2 \tau_7 \sum^{N_f} \int d^{8}\xi \, e^\Phi \Big[ 
\sqrt{-G_{8}^{(2)}} \Big]  \ ,
\end{eqnarray}
Here $\kappa_{10}$ is related to the ten dimensional Newton's constant $G_N$ {\it via}
\begin{eqnarray}
2 \kappa_{10}^2 = 16 \pi G_N = (2 \pi)^7 \alpha'^4 g_s^2 \ .
\end{eqnarray}
In the supergravity part of the action, $R$ is the Ricci scalar, $G_{10}$ is the background metric in Einstein frame, $\Phi$ is the dilaton and the five form field strength $F_5$ satisfies the self-duality condition: $\star F_5 = F_5$.

In the DBI part of the action $G_8^{(1,2)}$ is the induced metric on the seven brane and we have defined $g_s = e^{\Phi_*}$, where $\Phi_*$ is the asymptotic value of the dilaton field.\footnote{This implies that we need to impose the boundary condition $\Phi \to 0$ when we obtain the solutions.} With this definition we have 
\begin{eqnarray}
\tau_7  = \frac{1}{g_s} (2\pi)^{-7} \alpha'^4 \ 
\end{eqnarray}
such that 
\begin{eqnarray}
\frac{1}{2\kappa_{10}^2} = \frac{\tau_7}{g_s} = \frac{1}{(2\pi)^7 \alpha'^4 g_s^2} \ .
\end{eqnarray}
The factor of $2$ in front of the DBI piece comes from the fact that we are adding $N_f$ D7 and anti-D7 branes and they contribute equally to the action. The relative factors of $g_s$ can be understood as follows: the supergravity part of the action comes from the closed string sector and thus contains an overall factor of $g_s^{-2}$ whereas the DBI piece emerges from the open string sector and therefore contains an overall factor of $g_s^{-1}$. Note that here we will consider the back-reaction of the probes which are described by $\theta_i = \pi/2$ and $\phi_i = {\rm const}$ with $i = 1$ or $2$.

Applying the smearing procedure in the transverse directions leads to 
\begin{align}
\sum^{N_f} \int d^{8}\xi\, e^\Phi 
\sqrt{-G_8^{(1,2)}} \quad \to \quad &\frac{N_f}{4\pi}\int d^{10}x \, 
e^\Phi \,\sin\theta_1 \sqrt{-G_8^{(1,2)}} \ ,
\end{align}
where $G_8^{(1, 2)}$ represents the induced metric on the probe whose profile function is given by $\theta_{2, 1} = \pi/2 $ and $\phi_{2, 1} = {\rm const}$. The resulting ten dimensional action is given by 
\begin{eqnarray}
S = \frac{1}{2\kappa_{10}^2} \int d^{10}x \, \sqrt{-G_{10}} 
\Big[ R - \frac{1}{2} \partial_M \Phi \partial^M \Phi -\frac{1}{4} |F_5|^2 \Big]  & - & \frac{2 \tau_7 N_f}{4\pi}\int d^{10}x \, 
e^\Phi \,\sin\theta_1 \sqrt{-G_8^{(1)}} \nonumber \\
& - & \frac{2 \tau_7 N_f}{4\pi}\int d^{10}x \, 
e^\Phi \,\sin\theta_2 \sqrt{-G_8^{(2)}} \ . \nonumber\\
\end{eqnarray}
The equations of motion read
\begin{align}
E_{MN} &= T^{(1)}_{MN}+ T^{(2)}_{MN}+ T^{(3)}_{MN} \ , \\
D^M \partial_M \Phi &=  \frac{2\kappa_{10}^2 ( 2 \tau_7 )}{\sqrt{-G_{10}}} \frac{N_f}{4\pi} e^\Phi 
\left(  \sin\theta_1 \sqrt{-G_8^{(1)}} + \sin\theta_2 \sqrt{-G_8^{(2)}} \right) \ , \\
dF_5 &= 0 \;,
\end{align}
where
\begin{eqnarray}
E_{MN} & = & R_{MN}-\frac{1}{2}R G_{MN} \ ,\\
T^{(1)}_{MN} & = & \frac{1}{2} \Big( \partial_M \Phi \partial_N \Phi -\frac{1}{2} G_{MN} 
\partial_P \Phi \partial^P \Phi \Big)  \ ,\\
T^{(2)}_{MN} & = & \frac{1}{96}F_{MPQRS}F_N^{~PQRS}-\frac{1}{8 \times 5!}F_{PQRST}F^{PQRST}G_{MN} \ ,\\
T^{(3)MN} & = & \frac{2\kappa_{10}^2}{\sqrt{-G_{10}}} 
\frac{\delta S_{\rm{flavor}}}{\delta G_{MN}} = -\frac{1}{2}\frac{ (2 N_f) g_s^*}{4\pi} \frac{e^\Phi}{\sqrt{-G_{10}}} 
\sum_{i=1,2} \sin \theta_i \sqrt{-G_8^{(i)}} 
G_8^{(i)ab} \delta_a^M \delta_b^N \ . 
\end{eqnarray}
We have now denoted the string coupling constant, $g_s \equiv g_s^*$. The superscript will be a helpful bookkeeping device that there will be a cut-off radial position, denoted by $r=r_*$, where the boundary theory will be defined in the back-reacted background. When the back-reaction vanishes, this cut-off surface $r_*$ can be taken to infinity.

In order to solve the equations of motion, we will work with the following ansatz (in Einstein frame), which has the full $SU(2)\times SU(2) \times U(1) \times {\mathbb Z}_2$ invariance 
\begin{align}
ds_{10}^2 &= h^{-1/2}(r)\left[-b(r) dt^2 + dx_i dx^i\right]+
h^{1/2}(r) \bigg\{\frac{dr^2}{b(r)} +
\frac{e^{2g(r)}}{6} \sum_{i=1,2} \left( d\theta_i^2 + \sin^2 \theta_i \, d\phi_i^2 \right)  \nonumber \\
& + \frac{e^{2f(r)}}{9} \left(d\psi + \sum_{i=1,2} \cos\theta_i \, d\phi_i \right)^2 \bigg\} \ , \\
F_5 &= k(r)h(r)^{3/4}\left(e^t \wedge e^{x^1} \wedge e^{x^2} \wedge e^{x^3} \wedge e^r + e^{\psi} \wedge e^{\theta_1} \wedge e^{\phi_1} \wedge e^{\theta_2} \wedge e^{\phi_2}\right) \ ,
\end{align}
where the vielbeins are defined as
\begin{eqnarray}
&& e^t = h^{-1/4} b^{1/2} dt \ , \quad e^{x^i} =  h^{-1/4} dx^i \ , \quad e^r = h^{1/4} b^{-1/2} dr \ , \\
&& e^{\psi} = \frac{1}{3} h^{1/4} e^f \left( d\psi + \cos\theta_1 d\phi_1 + \cos\theta_2 d \phi_2 \right)  \ , \\
&& e^{\theta_{1,2}} =  \frac{1}{\sqrt{6}} h^{1/4} e^g d\theta_{1,2} \ ,  \quad e^{\phi_{1,2}} = \frac{1}{\sqrt{6}} h^{1/4} e^g \sin\theta_{1,2} d\phi_{1,2} \ .
\end{eqnarray}
Here the functions $h(r)$, $b(r)$, $f(r)$, $g(r)$ and $\Phi(r)$ are unknown functions which we have to obtain by solving the equations of motion. The Bianchi identity $dF_5 = 0$ is trivially satisfied and the quantization condition 
\begin{equation}
 \frac{1}{2\kappa_{10}^2} \int F_{p+2}=N_c \tau_p \quad {\rm for} \quad p=3 
\end{equation}
yields
\begin{equation}
k(r) h(r)^2 e^{4g(r)+f(r)}=27 \pi g_s^* N_c l_s^4  = 4 L^4 : = \frac{27}{4} \lambda l_s^4 \ ,
\end{equation}
where the 't Hooft coupling $\lambda$ is defined as $\lambda = 4 \pi g_s^* N_c$. In deriving the above relations we have also used the fact that
\begin{eqnarray}
\tau_p = \frac{1}{g_s^{*}} (2\pi)^{-p} \alpha'^{- (p+1)/2} \ , 
\end{eqnarray}
where $\alpha'$ is the string tension. Thus the function $k(r)$ is determined in terms of $h(r),g(r)$ and $f(r)$. Moreover,
\begin{eqnarray}
\sqrt{-G_8^{(1,2)}} & = & \frac{1}{18} \sin \left(\theta _{2,1}\right) e^{f(r)+2 g(r)} \ , \\
\sqrt{-G_{10}} & = & \frac{1}{108} \sqrt{h(r)} \sin \theta _1 \sin \theta _2 e^{f(r)+4 g(r)} \ .
\end{eqnarray}
The resulting equations of motion are\footnote{These equations of motion are obtained by taking various linear combinations of the Einstein equations and the Klein-Gordon equation.}
\begin{align}
& \Phi ''(r) + \Phi '(r) \left(f'(r)+4g'(r)+\frac{b'(r)}{b(r)}\right) = \frac{3 (2 N_f) g_s^*}{\pi b(r)}e^{\Phi(r) -2g(r)} \ , \label{eqns} \\
& 4 \left(b'(r) g'(r)+b(r) \left(f'(r) g'(r)+g''(r)+4 g'(r)^2\right)+2 e^{2 f(r)-4 g(r)}-6 e^{-2 g(r)}\right)+\frac{3 (2 N_f) g_s^* e^{\Phi (r)-2 g(r)}}{\pi } = 0 \ , \\
& 2 b'(r) f'(r)+\frac{b'(r) h'(r)}{h(r)}+8 b(r) f'(r) g'(r)+\frac{16 L^8 e^{-2 (f(r)+4 g(r))}-b(r) h'(r)^2}{h(r)^2}-8 b(r) g''(r) \nonumber\\
& -8 b(r) g'(r)^2-b(r) \Phi '(r)^2-8 e^{2 f(r)-4 g(r)} = 0 \ , \\
& b''(r)+b'(r) \left(f'(r)+4 g'(r)\right)=0 \ , \\
& f'(r) h'(r)+h(r) \left(2 f''(r)+2 f'(r)^2+8 g''(r)+8 g'(r)^2+\Phi '(r)^2\right)+4 g'(r) h'(r)+h''(r) = 0 \ , \\
& \frac{b'(r) f'(r)-4 e^{2 f(r)-4 g(r)}}{b(r)}+f''(r)+4 f'(r) g'(r)+f'(r)^2 = 0 \ . \label{eqne}
\end{align}
In the absence of any back-reaction ({\it i.e.}~setting $N_f = 0$) we will recover the usual AdS- Schwarzschild $\times T^{1,1}$ background given by
\begin{eqnarray}
&& h(r) = \frac{L^4}{r^4} \  , \quad b(r) = 1 - \frac{r_H^4}{r^4} \ , \quad \Phi = 0 \ , \\
&& e^f = e^g = r \ .
\end{eqnarray}

In terms of the redefined radial coordinate $\rho$, defined by
\begin{equation}
 e^{-f-4g}dr
= d\rho  \quad \Rightarrow \quad e^{f+4g} \frac{d}{dr} = \frac{d}{d\rho} 
\end{equation}
the equations of motion take the following simpler form
\begin{align}
& b''(\rho) = 0 \label{eq:b} \ , \\
& \Phi ''(\rho)+\frac{1}{b(\rho)}\Phi '(\rho)b'(\rho) =  \frac{3 (2 N_f) g_s^*}{\pi b(\rho)}e^{\Phi(\rho)+2f(\rho) +6g(\rho)} \ , \\
& f ''(\rho)+\frac{1}{b(\rho)}f'(\rho)b'(\rho) = \frac{4}{b(\rho)} e^{4f(\rho) +4g(\rho)} \ , \\
& g ''(\rho)+\frac{1}{b(\rho)}g'(\rho)b'(\rho) = \frac{1}{b(\rho)}\left[-2 e^{4f(\rho) +4g(\rho)}+6 e^{2f(\rho) +6g(\rho)}-\frac{3 (2 N_f) g_s^*}{4 \pi}e^{\Phi(\rho)+2f(\rho) +6g(\rho)}\right] \ , \\
& h(\rho)h''(\rho)-h'(\rho)^2+\frac{h(\rho)}{b(\rho)}h'(\rho)b'(\rho)+\frac{16 L^8}{b(\rho)} = 0 \ , \label{eq:h}
\end{align}
with an additional constraint equation,
\begin{align}
&\frac{h''(\rho)}{h(\rho)}+\Phi '(\rho)^2-24 g'(\rho)^2- 16g'(\rho)f'(\rho)+ 2 f ''(\rho)+8 g ''(\rho) = 0 \ . \label{constrainteqn}
\end{align}
Note that the equation for $b(\rho)$ completely decouples from the rest and is insensitive to the back-reaction. We can solve (\ref{eq:b}) and (\ref{eq:h}) analytically, yielding
\begin{align}
& b(\rho) = b_0 \left(1 -\frac{\rho}{\rho_H}\right) \ , \label{solbrho} \\
& h(\rho) = \pm \frac{8 L^4 \sqrt{b(\rho)}}{h_0} \cosh \left[ \frac{h_1 - \log b(\rho)}{2 b'(\rho)} h_0 \right]  \ . \label{solh}
\end{align}
The constant $\rho_H$ denotes the location of the event-horizon. Also, $b_0$, $h_0$ and $h_1$ are three undetermined constants.

\renewcommand{\theequation}{C.\arabic{equation}}
\setcounter{equation}{0}  
\section*{Appendix C. The perturbative solution}
\addcontentsline{toc}{section}{Appendix C. The perturbative solution}

Recall that the equation of motion for $b(\rho)$ completely decouples from the rest of the equations and can be solved exactly. We can find perturbative solutions for $\Phi(\rho)$, $f(\rho)$, $h(\rho)$ and $g(\rho)$ by employing the following expansion
\begin{eqnarray}
\Phi(\rho)=\Phi_0 + \epsilon \Phi_1(\rho) \ , \\
f(\rho)=f_0(\rho)+\epsilon f_1(\rho)\ , \\
g(\rho)=f_0(\rho)+\epsilon g_1(\rho) \ , \\
h(\rho)=h_0(\rho)+\epsilon h_1(\rho) \ ,
\end{eqnarray} 
where $\Phi_0$, $h_0(\rho)$ and $f_0(\rho)$ represent the KW-background solution and it should be noted that $e^{f_0}=e^{g_0}$ for the KW solution, given by
\begin{equation}
h_0(\rho(r)) = \frac{L^4}{r^4} \ , \quad \Phi_0 = 0 \ , \quad e^{f_0(\rho(r))} = r \ .
\end{equation}
Here we define 
\begin{equation}
\epsilon:=\frac{6}{\pi} g_s^{*} N_f = \frac{3}{2 \pi^2} \left(\frac{\lambda N_f}{N_c}\right) \ .
\end{equation}
The redefined radial coordinate $\rho$ is given by
\begin{equation}
\rho=-\frac{1}{4 r^4} + O(\epsilon)
\end{equation}
and $h_0(\rho)$, $f_0(\rho)$ are given by 
\begin{align}
f_0(\rho)&=-\frac{1}{4}\ln(- 4 \rho) \ ,\\
h_0(\rho)&=-4 R^4 \rho \ . 
\end{align}
Using equation (\ref{solb}), $b(\rho)$ is given by
\begin{align}
&b(\rho) = \left(1 -\frac{\rho}{\rho_H}\right) \ .
\end{align}
We have set $b_0=1$ since this parameter just scales the time-direction. Therefore, at the first order in $\epsilon$, we obtain the following equations
\begin{align} 
\Phi_1 ''(\rho)+\frac{1}{b(\rho)}\Phi_1 '(\rho)b'(\rho)=&\frac{1}{16 \rho^2 b(\rho)} \ ,\label{eqnphi1}\\
f_1 ''(\rho)+\frac{1}{b(\rho)}f_1'(\rho)b'(\rho)=& \frac{1}{\rho^2 b(\rho)}\left(f_1(\rho)+g_1(\rho)\right) \ , \label{eqnf1}\\
g_1 ''(\rho)+\frac{1}{b(\rho)}g_1'(\rho)b'(\rho)=& \frac{1}{4 \rho^2 b(\rho)}\left(f_1(\rho)+7 g_1(\rho)-\frac{1}{16}\right) \ , \label{eqng1}\\
\rho (\rho_H-\rho) h_1''(\rho)+(\rho-2\rho_H) & h_1'(\rho)-h_1(\rho)=0 \ . \label{eqnh1}
\end{align}
Equations (\ref{eqnphi1}, \ref{eqnh1}) can be solved analytically, yielding
\begin{align}
\Phi_1(\rho)=& c_1 -\frac{1}{16}\ln\left(\frac{\rho}{\rho_*}\right) - c_2\ln\left(1-\frac{\rho}{\rho_H}\right) \ ,\\
h_1(\rho)=& - c_3 \left(\frac{\rho}{\rho_H}\right) + 2(c_3+2c_4) +c_4 \left(2-\frac{\rho}{\rho_H}\right)\ln \left(1-\frac{\rho}{\rho_H}\right) \ .\label{solh1}
\end{align}
Now imposing the boundary condition $\Phi_1(\rho=\rho_*)=0$ will set $c_1 = 0$, where $\rho_*$ represents an UV cutoff. The constants $c_2$, $c_3$ and $c_4$ are still undetermined. The constraint equation (\ref{constrainteqn}), to the first order in $\epsilon$, becomes:
\begin{equation}
2(f_1 ''(\rho)+4 g_1 ''(\rho))+ \frac{h_1 ''(\rho)}{h_0 (\rho)}- 16 f_0 '(\rho)(f_1 '(\rho)+4 g_1 '(\rho))=0 \ .
\end{equation}
Using equation (\ref{solh1}), we get
\begin{equation}
2(f_1 ''(\rho)+4 g_1 ''(\rho))+ \frac{4}{\rho}(f_1 '(\rho)+4 g_1 '(\rho))=-\frac{c_4}{4 L^4 \left(\rho -\rho _H\right)^2 \rho _H} \ .
\end{equation}
The last equation can be solved analytically to give:
\begin{equation}
f_1(\rho)+4 g_1(\rho) = c_6 + \frac{c_5}{\rho L^4} + \frac{c_4\left(1-\frac{\rho}{2\rho_H}\right) \ln \left(1-\frac{\rho}{\rho_H}\right)}{4  \rho L^4} \ . \label{coneqn}
\end{equation}
Eqs.~(\ref{eqnf1}, \ref{eqng1}) can be rewritten as
\begin{align}
f_1 ''(\rho)+\frac{f_1'(\rho)}{\rho -\rho_H}=& \frac{\rho_H}{\rho^2 \rho_H - \rho^3}\left(f_1(\rho)+g_1(\rho)\right) \ , \\
g_1 ''(\rho)+\frac{g_1'(\rho)}{\rho-\rho_H}=& \frac{\rho_H}{\rho^2 \rho_H - \rho^3}\left(\frac{1}{4}f_1(\rho)+\frac{7}{4} g_1(\rho)-\frac{1}{64}\right) \ .
\end{align}
Combining these two equations, we get
\begin{equation}
\left(f_1''(\rho)+4 g_1''(\rho)\right)+\frac{1}{\rho-\rho_H}\left(f_1'(\rho)+4g_1'(\rho)\right)=\frac{\rho_H}{\rho^2 \rho_H - \rho^3} \left[2\left(f_1(\rho)+4g_1(\rho)\right)-\frac{1}{16}\right] \ .
\end{equation}
Now using equation (\ref{coneqn}), we get the following condition:
\begin{align}
 c_6=\frac{1}{32}\left[\frac{8(c_4 - 2 c_5)}{ L^4 \rho _H}+1\right]
\end{align}
and $f_1(\rho)$ can be written as:
\begin{equation}
f_1(\rho)=\frac{1}{32} \left(\frac{8 (c_4-2c_5)}{L^4 \rho _H}+1\right)+\frac{c_4 \left(1-\frac{\rho }{2 \rho _H}\right) \ln \left(1-\frac{\rho }{\rho _H}\right)}{4 \rho  L^4}+\frac{c_5}{\rho  L^4}-4 g_1(\rho ) \ .
\end{equation}
With that we have the following equation for $g_1(\rho)$:
\begin{align}
4 g_1''(\rho )+\frac{4 g_1'(\rho )}{\rho -\rho _H}+\frac{3 g_1(\rho ) \rho _H}{\rho ^2 \left(\rho -\rho _H\right)}  = & \frac{\rho _H}{32 \rho ^2 \left(\rho -\rho _H\right)}-\frac{c_4 \rho _H \log \left(1-\frac{\rho }{\rho _H}\right)}{4 \rho ^3 L^4 \left(\rho -\rho _H\right)}\nonumber\\
& -  \frac{c_4}{4 \rho ^2 L^4 \left(\rho -\rho _H\right)}+\frac{c_4 \log \left(1-\frac{\rho }{\rho _H}\right)}{8 \rho ^2 L^4 \left(\rho -\rho _H\right)} \nonumber\\ 
& -  \frac{c_5 \rho _H}{\rho ^3 L^4 \left(\rho -\rho _H\right)}+\frac{c_5}{2 \rho ^2 L^4 \left(\rho -\rho _H\right)} \ .
\end{align}

\noindent {\bf Boundary conditions:}

Before solving for $g_1(\rho)$, we can determine some of the constants by demanding regularity of the Ricci scalar on the horizon. Using the equations of motion, the Ricci scalar $S$ is given by
\begin{equation}
R := R_{MN}G^{MN}=\frac{b(\rho)}{h(\rho)^{1/2}}e^{-2(f(\rho)+4g(\rho))}\left[\frac{6 N_f g_s^*}{\pi b(\rho)}e^{\Phi(\rho)+2f(\rho) +6g(\rho)}+\frac{1}{2}\Phi '(\rho)^2\right] \ .
\end{equation}
Now the condition that $S$ is regular at the horizon tells us:
\begin{equation}
c_4=0 \ , \quad c_2=0 \ .
\end{equation}
Note that the condition $c_4 = 0$ could have also been obtained from demanding regularity of $f_1(\rho)$ at the horizon. We can also choose $c_1=0$. After redefining the constant $c_5$ ($\frac{c_5}{L^4 \rho_H}=c_5$), we have
\begin{align}
&b(\rho) = \left(1 -\frac{\rho}{\rho_H}\right) \ , \\
& \Phi_1(\rho)= -\frac{1}{16}\ln\left(\frac{\rho} {\rho_*}\right) \ , \\
& h_1(\rho)= c_3 \left(2-\frac{\rho}{\rho_H}\right) \ , \\
& f_1(\rho)=\frac{1}{32} \left(-16 c_5+1\right)+\frac{c_5 \rho_H}{\rho }-4 g_1(\rho ) \ , \label{equationf1}
\end{align}
with $ g_1(\rho)$ obeying
\begin{align}
4 g_1''(\rho )+\frac{4 g_1'(\rho )}{\rho -\rho _H}+\frac{3 g_1(\rho ) \rho _H}{\rho ^2 \left(\rho -\rho _H\right)}= & \frac{\rho _H}{32 \rho ^2 \left(\rho -\rho _H\right)}-\frac{c_5 \rho _H^2}{\rho ^3  \left(\rho -\rho _H\right)}+\frac{c_5 \rho_H}{2 \rho ^2  \left(\rho -\rho _H\right)} \ .
\end{align}
The general solution of this equation is given by
\begin{align}\label{solutiong1}
g_1(\rho)=\frac{1}{96}+\frac{c_5 \rho_H}{5 \rho}-\frac{c_5}{10}+\frac{c_7 \rho ^{3/2} \, _2F_1\left(\frac{3}{2},\frac{3}{2};3;\frac{\rho }{\rho_H}\right)}{\rho _H{}^{3/2}}+\frac{c_8 \left(2 E\left(1-\frac{\rho }{\rho _H}\right)-\frac{\rho  K\left(1-\frac{\rho }{\rho _H}\right)}{\rho _H}\right)}{\sqrt{\rho/\rho_H }} \ ,
\end{align}
where, $E(x)$ and $K(x)$ are the Elliptic functions. We need to impose $c_7=0$ for $g_1(\rho)$ to be regular at the horizon. Therefore, using equations (\ref{equationf1},\ref{solutiong1}), we get
\begin{align}
g_1(\rho)=& \frac{1}{96}+\frac{c_5 \rho_H}{5 \rho}-\frac{c_5}{10}+\frac{c_8 \left(2 E\left(1-\frac{\rho }{\rho _H}\right)-\frac{\rho  K\left(1-\frac{\rho }{\rho _H}\right)}{\rho _H}\right)}{\sqrt{\rho/\rho_H }} \ , \label{solg1} \\
f_1(\rho)=& -\frac{1}{96}+\frac{c_5 \rho_H}{5 \rho}-\frac{c_5}{10}+\frac{4 c_8 \sqrt{\rho } K\left(1-\frac{\rho }{\rho _H}\right)}{\sqrt{\rho _H}}-\frac{8 c_8 \sqrt{\rho _H} E\left(1-\frac{\rho }{\rho _H}\right)}{\sqrt{\rho }} \ . \label{solf1}
\end{align}
Now using equation (\ref{equationf1}) and the fact that $dr=\exp(f(\rho)+4g(\rho))d\rho$, to the first order in $\epsilon$ we have
\begin{equation}
r(\rho)=\frac{1}{\sqrt{2} (-\rho)^{1/4}}\left[1+ \epsilon \left(\frac{1}{32}- \frac{c_5}{2}+\frac{c_5 \rho_H}{5 \rho}\right)\right] \ , \label{eqnr}
\end{equation} 
and $r_H$ is given by
\begin{equation}
r_H=\frac{1}{\sqrt{2} (-\rho_H)^{1/4}}\left[1+ \epsilon \left(\frac{1}{32}- \frac{3 c_5}{10}\right)\right] \ .
\end{equation} 
Equation (\ref{eqnr}) can be inverted easily to get $\rho(r)$
\begin{equation}
\rho(r)=-\frac{1}{4 r^4}-\epsilon\left[\frac{1}{ r^4}\left(\frac{1}{32}- \frac{c_5}{2}\right)-\frac{4 c_5 \rho_H}{5 }\right]+ \cO(\epsilon^2) \ .
\end{equation}

Finally, we have the first order solutions in terms of the $r$ coordinate:
\begin{align}
&b(r) =\left(1-\frac{r_H^4}{r^4}\right)\left(1-\frac{4}{5}\epsilon c_5\right) \ , \\
& \Phi(r)= \frac{\epsilon}{4}\ln\left(\frac{r}{r_*}\right) \ , \\
& h(r)=\frac{L^4}{r^4}+\epsilon\left[\frac{L^4}{r^4}\left(\frac{1}{8}-2c_5\right)+ \frac{4 L^4 c_5}{5 r_H^4}+c_3\left(2-\frac{r_H^4}{r^4}\right) \right] \ , \\
& e^{f(r)}=r\left[1+\epsilon\left(\frac{2 c_5}{5}-\frac{1}{24}+4 c_8  \left(\frac{r_H^2}{r^2}\right)K\left(1-\frac{r_H^4}{r^4}\right)- 8 c_8  \left(\frac{r^2}{r_H^2}\right)E\left(1-\frac{r_H^4}{r^4}\right)\right)\right]  \ , \\
& e^{g(r)}=r\left[1+\epsilon\left(\frac{2 c_5}{5}-\frac{1}{48}- c_8  \left(\frac{r_H^2}{r^2}\right)K\left(1-\frac{r_H^4}{r^4}\right)+2 c_8  \left(\frac{r^2}{r_H^2}\right)E\left(1-\frac{r_H^4}{r^4}\right)\right)\right]  \  .
\end{align}
After absorbing the constant factor $\left(1-\frac{4}{5}\epsilon c_5\right)$ of $b(r)$ in all the other functions and then redefining the $x^i$ coordinates, the background solution can be given by
\begin{align}
&b(r) =\left(1-\frac{r_H^4}{r^4}\right) \ , \label{solgenb} \\
& \Phi(r)= \frac{\epsilon}{4}\ln\left(\frac{r}{r_*}\right) \ , \\
& h(r)=\frac{L^4}{r^4}+\epsilon\left[\frac{L^4}{r^4}\left(\frac{1}{8}-\frac{2}{5}c_5\right)+ \frac{4 L^4 c_5}{5 r_H^4}+c_3\left(2-\frac{r_H^4}{r^4}\right) \right] \ , \\
& e^{f(r)}=r\left[1+\epsilon\left(-\frac{1}{24}+4 c_8  \left(\frac{r_H^2}{r^2}\right)K\left(1-\frac{r_H^4}{r^4}\right)- 8 c_8  \left(\frac{r^2}{r_H^2}\right)E\left(1-\frac{r_H^4}{r^4}\right)\right)\right] \ , \\
& e^{g(r)}=r\left[1+\epsilon\left(-\frac{1}{48}- c_8  \left(\frac{r_H^2}{r^2}\right)K\left(1-\frac{r_H^4}{r^4}\right)+2 c_8  \left(\frac{r^2}{r_H^2}\right)E\left(1-\frac{r_H^4}{r^4}\right)\right)\right] \ . \label{solgeng}
\end{align}
The solution presented in (\ref{solgenb})-(\ref{solgeng}) is the most general solution of the equations of motion, which contains three undetermined constants denoted by $c_3$, $c_5$ and $c_8$ respectively. Below we will fix two of these constants and in the main text we will consider the background which eventually will have only one independent constant.

In general, the solution for $h(r)$ takes a form which is deformed compared to the warp factor which corresponds to pure AdS. Imposing the condition that $h(r) \sim L^4/r^4$ we get
\begin{equation} \label{c3c5choice}
c_3=-\frac{2 L^4 c_5}{5 r_H^4} \ .
\end{equation}
Alternatively we could choose both $c_3$ and $c_5$ to be zero to begin with. This choice of constants, in the dual field theory, corresponds to setting a dimension $8$ operator to zero. Thus we have fixed all constants of integration except $c_8$.

Now, it is important to note that the constant $c_8$ can be a function of $r_H$. In the limit $r\rightarrow r_*$, 
\begin{eqnarray}
e^{2g(r)} \ ,  e^{2 f(r)} \sim r^2 + \cO(1) \left( \epsilon c_8 \right) \frac{r_*^4}{r_H^2} \ ,
\end{eqnarray}
where $r_* \gg r_H$. But far away from the horizon, the metric should be independent of $r_H$. Therefore, $c_8$ is given by
\begin{equation}
c_8= r_H^2 \mathfrak{m}^{-2}  \  .
\end{equation}
The constant $\mathfrak{m}^{-2}$ does not depend on $r_H$ and it has the dimension of $\text{length}^{-2}$. Finally we have
\begin{align}
&b(r) =\left(1-\frac{r_H^4}{r^4}\right) \ ,\label{sols} \\
& \Phi(r)=\frac{\epsilon}{4}\ln\left(\frac{r}{r_*}\right) \ , \\
& h(r)=\frac{L^4}{r^4}\left(1+\frac{\epsilon}{8}\right) \ ,\\
& e^{f(r)}=r\left[1+\epsilon\left(-\frac{1}{24}+4 \mathfrak{m}^{-2}  \left(\frac{r_H^4}{r^2}\right)K\left(1-\frac{r_H^4}{r^4}\right)- 8 \mathfrak{m}^{-2} r^2 E\left(1-\frac{r_H^4}{r^4}\right)\right)\right] \ , \\
& e^{g(r)}=r\left[1+\epsilon\left(-\frac{1}{48}- \mathfrak{m}^{-2}  \left(\frac{r_H^4}{r^2}\right)K\left(1-\frac{r_H^4}{r^4}\right)+2 \mathfrak{m}^{-2} r^2 E\left(1-\frac{r_H^4}{r^4}\right)\right)\right] \ . \label{sole}
\end{align}
%

\renewcommand{\theequation}{D.\arabic{equation}}
\setcounter{equation}{0}  
\section*{Appendix D. Fluctuations of the classical profile}
\addcontentsline{toc}{section}{Appendix D. Fluctuations of the classical profile}

Let us denote the embedding of the probe brane by 
\begin{eqnarray} \label{flucx}
X^\mu = X_0^a \delta_a^\mu + \Sigma^i \delta_i^\mu \ , 
\end{eqnarray}
where $X_0^a$ denotes the classical profile of the probe, and $\Sigma^i$ denotes the transverse fluctuations. The indices $\mu, \nu = 0, \ldots 9$; the indices $i, j, k = 8, 9$ and finally we will make use of the indices $a, b, c = 0 \ldots 7$. Clearly, the indices $\mu, \nu$ {\it etc.} are used for the ten dimensional background, the indices $i, j, k$ are used to represent the transverse fluctuations (which in our coordinate system are denoted by $\delta\theta_1$ and $\delta\phi_1$ respectively), and finally the indices $a, b, c$ are used to represent the world volume of the probe brane.

The background metric $G_{\mu\nu}$ can be expanded under (\ref{flucx}) as
\begin{eqnarray}
G_{\mu\nu} = G_{\mu\nu}^0 + \partial_\rho G_{\mu\nu}^0 \Sigma^i \delta_i^\rho + \frac{1}{2} \partial_\rho\partial_\sigma G_{\mu\nu}^0 \Sigma^i \Sigma^j \delta_i^\rho \delta_j^\sigma + \ldots \ .
\end{eqnarray}
Here $G_{\mu\nu}^0$ denotes the background metric evaluated at the classical profile of the probe.

Now, the induced metric (up to quadratic order in the transverse fluctuations) on the probe D-brane can be obtained from the following formula
\begin{eqnarray}
g_{ab} & = & G_{\mu\nu} \partial_a X^\mu \partial_b X^\nu \nonumber\\
& = & g_{ab}^0 + g_{ab}^1 + g_{ab}^2 \ ,
\end{eqnarray}
where
\begin{eqnarray} \label{gab0}
g_{ab}^0 = G_{\mu\nu}^0 \partial_a X_0^{a'} \partial_b X_0^{b'} \delta_{a'}^\mu \delta_{b'}^\nu = G_{\mu\nu}^0 \partial_a X_0^\mu \partial_b X_0^\nu\ ,
\end{eqnarray}
\begin{eqnarray} \label{gab1}
g_{ab}^1 = \partial_\rho G_{\mu\nu}^0 \partial_a X_0^\mu \partial_b X_0^\nu \Sigma^i \delta_i^\rho + G_{\mu\nu}^0 \left[ \partial_a X_0^\mu \partial_b\Sigma^i\delta_i^\nu + \partial_b X_0^\nu \partial_a\Sigma^i\delta_i^\mu \right] \ ,
\end{eqnarray}
and 
\begin{eqnarray} \label{gab2}
g_{ab}^2 & = & G_{\mu\nu}^0 \partial_a \Sigma^i \partial_b \Sigma^j \delta_i^\mu \delta_j^\nu + \frac{1}{2} \partial_a X_0^\mu \partial_b X_0^\nu \partial_\rho \partial_\sigma G_{\mu\nu}^0 \Sigma^i \Sigma^j \delta_i^\rho \delta_j^\sigma \nonumber\\
& + & \partial_\rho G_{\mu\nu}^0 \Sigma^i \delta_i^\rho \left[ \partial_a X_0^\mu \partial_b \Sigma^j \delta_j^\nu + \partial_b X_0^\nu \partial_a \Sigma^j \delta_j^\mu \right] \ .
\end{eqnarray}
Now the action functional for the probe brane is given by
\begin{eqnarray}
S_{\rm probe} = - T_7 \left( \int d^8 \xi e^{\Phi} \sqrt{ - E_{ab}} - \int C_8 \right) \ ,
\end{eqnarray}
where $\xi$ represents the world volume coordinates, $C_8$ is an $8$-form potential and 
\begin{eqnarray}
E_{ab} = E_{ab}^0 + E_{ab}^1 + E_{ab}^2 \ , 
\end{eqnarray}
with
\begin{eqnarray}
E_{ab}^0 = g_{ab}^0 \ , \quad E_{ab}^1 = g_{ab}^1 + F_{ab} \ , \quad E_{ab}^2 = g_{ab}^2 \ ,
\end{eqnarray}
where $F_{ab}$ is the vector fluctuation.

Now we use the following fact
\begin{eqnarray}
{\rm det}(-E_{ab})^{1/2} = {\rm det}(-E_{ab}^0)^{1/2}  \left[ 1 + \frac{1}{2} {\rm Tr} M - \frac{1}{4} {\rm Tr} M^2 + \frac{1}{8} \left( {\rm Tr} M \right)^2\right] + \ldots \ ,
\end{eqnarray}
where
\begin{eqnarray}
M_a^c = g_0^{cb} E_{ba}^1 + g_0^{cb} E_{ba}^2 \ .
\end{eqnarray}

Now it is straightforward to check that at the quadratic order, we get the following contributions
\begin{eqnarray}
 \frac{1}{2} {\rm Tr} M \quad & \implies & \quad \frac{1}{2} g_0^{ab} g_{ab}^2 \ ,  \label{quad1} \\
 \frac{1}{8} \left({\rm Tr} M\right)^2 \quad & \implies & \quad \frac{1}{8} g_0^{ab} g_{ab}^1 g_0^{cd} g_{cd}^1 \ , \label{quad2} \\
 - \frac{1}{4} {\rm Tr} \left(M\right)^2 \quad & \implies & \quad  - \frac{1}{4} g_0^{cb}g_0^{a'b'} F_{ba'} F_{b'c}   - \frac{1}{4} g_0^{cb} g_0^{a'b'} g_{a'b}^1 g_{b'c}^1 \nonumber\\
 & - & \frac{1}{4} \left(g_0^{cb} g_0^{a'b'} - g_0^{ca'} g_0^{bb'} \right) g_{b'c}^1 F_{ba'} \ . \label{quad3}
\end{eqnarray}
Note that the above formulae are completely general for any background metric and any probe brane profile. So, in general, the vector and the scalar fluctuations will be coupled through the last term in (\ref{quad3}). On the other hand, the Wess-Zumino term gives the following contribution at the quadratic order
\begin{eqnarray}
\frac{1}{2} P[C_4] \wedge F \wedge F \ ,
\end{eqnarray}
where $P[C_4]$ denotes the pull-back of the background $C_4$ potential and $F$ is the vector fluctuation.

Now the background metric we have considered is of the following form,
\begin{eqnarray}
ds^2 = h^{-1/2} \left( - b dt^2 + d\vec{x}^2\right) & + & h^{1/2} \left( \frac{dr^2}{b} + \frac{e^{2g}}{6} \sum_{i=1,2} \left(d\theta_i^2 + \sin^2\theta_i d\phi_i^2 \right)\right. \nonumber\\
 & + & \left. \frac{e^{2f}}{9} \left(d\psi + \cos\theta_1 d\phi_1 + \cos\theta_2 d\phi_2 \right)^2 \right) \ .
\end{eqnarray}
The induced metric for the profile given by $\theta_1 = \pi/2$ and $\phi_1 = {\rm const}$ is obtained to be
\begin{eqnarray} \label{induced}
ds^2 = h^{-1/2} \left( - b dt^2 + d\vec{x}^2\right) & + & h^{1/2} \left( \frac{dr^2}{b} + \frac{e^{2g}}{6} \left(d\theta_2^2 + \sin^2\theta_2 d\phi_2^2 \right)\right. \nonumber\\
 & + & \left. \frac{e^{2f}}{9} \left(d\psi + \cos\theta_2 d\phi_2 \right)^2 \right) \ .
\end{eqnarray}
Using the above informations and the expressions in (\ref{gab1})-(\ref{gab2}) it can be shown that 
\begin{eqnarray} \label{gab1here}
g_{ab}^1 = G_{\psi\phi_1}^0 \left( \partial_a \delta \phi_1 \delta_b^\psi + \partial_b \delta \phi_1 \delta_a^\psi \right) = 0  \ .
\end{eqnarray}
The last equality follows from $G_{\psi\phi_1}^0 = \cos\theta_1^0 = \cos \left(\pi/2\right) = 0 $. And
\begin{eqnarray} \label{gab2here}
g_{ab}^2 & = & G_{\theta_1\theta_1}^0 \partial_a\delta\theta_1 \partial_b \delta\theta_1 + G_{\phi_1\phi_1}^0 \partial_a\delta\phi_1 \partial_b \delta\phi_1 \nonumber\\
& +& \partial_{\theta_1} G_{\phi_1\phi_2}^0 \delta\theta_1 \left( \partial_a \delta\phi_1 \delta_b^{\phi_2} + \partial_b \delta\phi_1 \delta_a^{\phi_2} \right) + \partial_{\theta_1} G_{\phi_1\psi}^0 \delta\theta_1 \left( \partial_a \delta\phi_1 \delta_b^{\psi} + \partial_b \delta\phi_1 \delta_a^{\psi} \right)
\end{eqnarray}

Further simplifications follow: using the contributions listed in (\ref{quad1})-(\ref{quad3}), it is also straightforward to check that for the induced metric in (\ref{induced})
\begin{eqnarray}
- \frac{1}{4} \left(g_0^{cb} g_0^{a'b'} - g_0^{ca'} g_0^{bb'} \right) g_{b'c}^1 F_{ba'} = 0 \ .
\end{eqnarray}
Hence the vector fluctuations decouple completely from the scalar ones. We will not worry about the vector fluctuations henceforth.

Finally, the Lagrangian for the scalar fluctuation is obtained to be
\begin{eqnarray} \label{quadscalar}
\cL & = &  e^{\Phi} \sqrt{- E_{ab}^0} \left[ \frac{1}{2} g_{0}^{ab} G_{\theta_1\theta_1}^0 \partial_a \delta \theta_1 \partial_b\delta \theta_1 + \frac{1}{2} g_0^{ab} G_{\phi_1\phi_1}^0  \partial_a\delta\phi_1 \partial_b \delta\phi_1 \right. \nonumber\\
& + &  \left(\partial_{\theta_1} G_{\phi_1\phi_2}^0\right) \left[ g_0^{\phi_2\phi_2} \partial_{\phi_2} \delta\phi_1 + g_0^{\psi\phi_2} \partial_\psi \delta \phi_1\right] \delta\theta_1 \nonumber\\
& + & \left. \left(\partial_{\theta_1} G_{\psi \phi_1}^0\right) \left[ g_0^{\psi\psi} \partial_{\psi} \delta\phi_1 + g_0^{\psi\phi_2} \partial_{\phi_2} \delta \phi_1\right] \delta\theta_1 \right] \ .
\end{eqnarray}
The equations of motion for $\delta\theta_1$ and $\delta\phi_1$ resulting from the Lagrangian in (\ref{quadscalar}) are given by
\begin{eqnarray} \label{fluctheta}
&& \partial_a \left[e^{\Phi} \sqrt{- E^0} g_0^{ab} G_{\theta_1\theta_1}^0 \partial_b \delta\theta_1\right] - \nonumber\\
&& e^{\Phi} \sqrt{- E^0} \left( \partial_{\theta_1} G_{\phi_1\phi_2}^0 g_0^{\phi_2 \phi_2} + \partial_{\theta_1} G_{\psi \phi_1}^0 g_0^{\phi_2 \psi} \right) \partial_{\phi_2} \delta\phi_1  \nonumber\\
&& - e^{\Phi} \sqrt{- E^0} \left( \partial_{\theta_1} G_{\psi \phi_1}^0 g_0^{\psi \psi} + \partial_{\theta_1} G_{\phi_1 \phi_2}^0 g_0^{\phi_2 \psi} \right) \partial_{\psi} \delta\phi_1  = 0 \ ,
\end{eqnarray}
and
\begin{eqnarray} \label{flucphi}
&& \partial_a \left[e^{\Phi} \sqrt{- E^0} g_0^{ab} G_{\phi_1\phi_1}^0 \partial_b \delta\phi_1\right]  \nonumber\\
&& + \partial_{\phi_2} \left[e^{\Phi} \sqrt{- E^0} \left( \partial_{\theta_1} G_{\phi_1\phi_2}^0 g_0^{\phi_2 \phi_2} + \partial_{\theta_1} G_{\psi \phi_1}^0 g_0^{\phi_2 \psi} \right)  \delta\theta_1\right]  \nonumber\\
&& + \partial_{\psi} \left[ e^{\Phi} \sqrt{- E^0} \left( \partial_{\theta_1} G_{\psi \phi_1}^0 g_0^{\psi \psi} + \partial_{\theta_1} G_{\phi_1 \phi_2}^0 g_0^{\phi_2 \psi} \right)  \delta\theta_1 \right] = 0 \ ,
\end{eqnarray}
It is clear from (\ref{fluctheta}) and (\ref{flucphi}) that in general the scalar fluctuations are coupled; specifically, they are coupled {\it via} the fluctuation modes along $\psi$ and $\phi_2$ directions. However, using the expressions for the background and the induced metric, it can be shown that the coupled equations take the simpler form
\begin{eqnarray} 
&& \partial_a \left[ e^{\Phi} \sqrt{-E^0} g_0^{ab} G_{\theta_1\theta_1}^0 \partial_b \delta\theta_1 \right] + e^{\Phi} \sqrt{-E^0} \partial_\psi \delta\phi_1 = 0 \ , \label{scalarfluc1} \\
&&  \partial_a \left[ e^{\Phi} \sqrt{-E^0} g_0^{ab} G_{\phi_1\phi_1}^0 \partial_b \delta\phi_1 \right] - e^{\Phi} \sqrt{-E^0} \partial_\psi \delta\theta_1 = 0 \ . \label{scalarfluc2}
\end{eqnarray}
For the above equations the method of separation of variables cannot be applied.



\end{document}